\documentclass[
 reprint,
 amsmath,amssymb,
 aps, 
pra,
]{revtex4-1}

\usepackage{xcolor}
\usepackage{dsfont}
\usepackage{graphicx}% Include figure files
\usepackage{dcolumn}% Align table columns on decimal point
\usepackage{bm}% bold math
\usepackage{verbatim}
\usepackage{listings}
\usepackage{bm,color,float,dcolumn,amssymb,amsmath}
\usepackage[a4paper, margin=2cm]{geometry}

\usepackage{color}
\usepackage[utf8]{inputenc}
\usepackage[english]{babel}

\usepackage{xr}
\definecolor{myblue}{rgb}{0.2,0.2,0.8}
\definecolor{myblack}{rgb}{0,0,0}
\definecolor{myurl}{rgb}{0.1,0.1,0.4}

\usepackage[colorlinks=true,citecolor=myblue,linkcolor=myblack,urlcolor=myurl]{hyperref}

\usepackage{braket}

\newcommand{\tr}{\mathrm{Tr}}

\newcommand{\eye}{\mathds{1}}

\newcommand{\Tr}{\text{Tr}}
\newcommand{\LL}{\mathcal{L}}

\newcommand{\D}{\text{d}}

\begin{document}

\title{Optimal  cycles for low-dissipation heat engines}

\author{Paolo Abiuso}
\email{paolo.abiuso@icfo.eu}
\affiliation{ICFO – Institut de Ciències Fotòniques, The Barcelona Institute of Science and Technology,08860 Castelldefels (Barcelona), Spain}

\author{Mart\' i Perarnau-Llobet}
\affiliation{Max-Planck-Institut für Quantenoptik, D-85748 Garching, Germany}

\date{\today}

\begin{abstract}
%In this article %we consider a class of machines whose performance hinges on the rate of their output. We
%we introduce and
%In this article 
We consider the  optimization of a finite-time Carnot engine characterized by small dissipations. 
We bound the power with a simple inequality and show that the optimal strategy is to perform small cycles around a given working point, which can be thus chosen optimally. Remarkably, this optimal  point is independent of the figure of merit combining power and efficiency that is being maximized. Furthermore, for a general class of dissipative dynamics the maximal power output  becomes proportional to the heat capacity of the working substance. Since the heat capacity can scale supra-extensively with the number of constituents of the engine, 
%(e.g. in a phase transition point)
this enables us to design optimal many-body Carnot engines reaching maximum efficiency at finite power per constituent in the thermodynamic limit.  %, which we argue to be optimal 
% inspired by previous proposals, while proving the optimality in the slow-driving regime.
\end{abstract}

\maketitle
%\section{Introduction}
The Carnot engine has a pivotal role in  thermodynamics, both from a fundamental and applied perspective, being %a building block of more complex thermal machines (e.g. refrigerators) and 
the reference point for other engines in terms of efficiency \cite{carnot1824reflections,callen1998thermodynamics}. It is thus of paramount importance to understand its limits and  strategies for its best utilization.
In this article, we consider the optimization of a finite-time Carnot cycle 
%between a hot and a cold thermal baths at temperature $T_c$ and $T_h$, respectively. We work  
within  the so called \emph{low-dissipation} (LD) regime~\cite{
esposito-EMP-bounds,Guo2013,VandenBroeck2013,Hernndez2015,Holubec2015,Holubec2016,ma2018universal,Tomas2012,deTomas2013,Wang2012,COP2013,slowdriving},  %where the dissipation is inversely proportional to the time of the process.
where the driving of the control parameters is slow but finite. 
 Previous studies of Carnot engines in the LD regime have considered  bounds on the reachable efficiencies~\cite{esposito-EMP-bounds}, 
%Within the LD regime, the efficiency at maximum power  $\eta^*$ of the  engine is known to  be bounded as $\eta_C/2 \leq \eta^*\leq  \eta_C/(2-\eta_C)$, where $\eta_C$ is the Carnot efficiency. Furthermore, when the engine operates symmetrically between the hot and cold reservoir (at temperature $T_h$ and $T_c$, respectively), $\eta*$ is given by the Curzon-Ahlborn efficiency~\cite{curzon-ahlborn},
%\begin{align}
%\label{etaCA}
%\eta_{CA} =1-\sqrt{\frac{T_c}{T_h}}. 
%\end{align}
tradeoffs between efficiency and power~\cite{Holubec2015,Holubec2016,Shiraishi2016,ma2018universal}, the coefficient of performance of refrigerators~\cite{Wang2012,COP2013},  the impact of the spectral density of the thermal baths~\cite{slowdriving}, and other thermodynamic figures of merit~\cite{Tomas2012,deTomas2013}. 
Despite this remarkable progress, the following crucial question has remained unaddressed: given a certain level of control on the working substance (e.g. some parameters of the Hamiltonian, or some macroscopic variables such as volume or pressure), what is the optimal cyclic modulation of the control parameters to maximise the power output (or, more generally, any figure of merit involving power and efficiency~\cite{Holubec2015,Holubec2016,Shiraishi2016,ma2018universal}) of a finite-time Carnot engine? Such an optimal cycle has been designed for a single-qubit engine in \cite{Esposito2010pre,abiuso_n-M}, but a general understanding is lacking.

Using recent insights on a geometrical approach to thermodynamics~\cite{Weinhold,Ruppeiner1979,Salamon1983a,Schlogl1985,Nulton1,Crooks2007,Sivak2012a,Bonanca2014d,Zulkowski2015,Mandal2016a,marti2018thermodynamiclength,martimatteo_fluctuations}, we show that, given any reasonable figure of merit involving power and efficiency,  the optimal control strategy is to perform infinitesimal cycles around a fixed point. Furthermore, when the thermalization of the relevant quantities  can be described by a single time-scale $\tau_{\rm eq}$ (see details below),  the optimal power output becomes proportional to $\mathcal{C}/\tau_{\rm eq}$, where $\mathcal{C}$ is the heat capacity of the working substance (WS). Hence, the  optimization of the heat engine cycle  becomes intimately related to the maximization of $\mathcal{C}$ of the WS % given a certain level of control
 (interestingly, maximizing $\mathcal{C}$ is also crucial in  thermometry~\cite{Correa-Mehboudi_optimalthermalprobes,hofer2017fundamental,Hovhannisyan2018,mehboudi2018thermometry}).  %This turns out to be a crucial observation, as it allows to naturally combine the optimal control problem with the physical nature of the working substance. 

We then use these insights to design many-body heat engines that can operate at Carnot efficiency with finite power per constituent of the WS through a supraextensive scaling of $\mathcal{C}/\tau_{\rm eq}$ (e.g. in a phase transition), in the spirit of~\cite{Finite_power_carnot_attainability,critical_engine} (see also \cite{critical_engine,critical_engine_fluctuations,Ma2017,Chand2018}).
Despite differences w.r.t. previous proposals, which were based on Otto engines \cite{Finite_power_carnot_attainability,critical_engine}, we find the same asymptotic scalings for performance, while proving by construction their optimality in the slow-driving regime. Other recent proposals towards the possibility of  reaching Carnot efficiency at finite power have been developed in 
%Furthermore, we also argue that these proposals can overcome the presence of macroscopic fluctuations~\cite{critical_engine_fluctuations} in case of critical speedup of thermalization. %, so that the relative fluctuations per unit of time also disappear in the thermodynamic limit. %In contrast to previous proposals of engines working (close to) Carnot efficiency \cite{}, 
%By construction, the derived cycle is also provably optimal within its regime of validity. 
%Hence we make a  step forward on our understanding of the possibility of (asymptotically) reaching Carnot efficiency at finite power, which has been recently suggested  in several contexts
~\cite{mintchev2013thermoelectric,critical_engine,Finite_power_carnot_attainability,
critical_engine_fluctuations,Polettini2015,Koning2016,Polettini2017,Johnson2018,Holubec2018,
erdman2018maximum} (see also \cite{Shiraishi2015,Shiraishi2016,Shiraishi2017} for no-go results %~\cite{Note1}).
 \footnote{Note that these results are not in contradiction with the present analysis, as we are concerned with reaching Carnot efficiency \emph{asymptotically}.}). % on the impossibility of reaching Carnot at finite power given rather mild assumptions \footnote{Note that these results are not in contradiction with the present analysis, as we are concerned with reaching Carnot efficiency \emph{asymptotically}.}).

\emph{Finite-time  Carnot cycle}. We consider a quantum working substance (WS) with a driven Hamiltonian  $H(t)$, which interacts alternatively with a cold ($B_c$) and a hot ($B_h$) heat bath at  temperature $T_c$ and $T_h$, respectively 
(the results presented in this article can be naturally extended to classical systems). The Carnot cycle  consists of four steps:
% two fast adiabatic processes (whose time can be neglected), and two slow isothermal processes where the engine interacts either with $B_{h}$ or $B_c$ for a time $\tau_h$  or $\tau_c$:
(i)~While being coupled to $B_c$, $H(t)$ is modified continuously from~$H(0)= H^{(X)}$ to~$H(\tau_c^-)= H^{(Y)}$  for a time~$\tau_c$. %, which leads to the release of heat.
(ii)~With the system isolated from the reservoirs, an adiabatic process  is performed taking~$H^{(Y)}\rightarrow H^{(Y)} T_h/T_c$. %This is a very fast process (compared to the relaxation timescales) and its time is neglected.
During this process $H(t)$ satisfies $[H(t),H(t')]=0$ $\forall t,t'$ and commutes with the initial state (a Gibbs state w.r.t~$H(\tau_c^-)$), so it is  possible to perform it arbitrarily quickly without affecting the state (hence corresponding to Hamiltonian quenches).
%\textcolor{red}{ Hence, in the following we assume that the adiabats are instantaneous, thus corresponding to Hamiltonian quenches.}
(iii)~While being coupled to $B_h$, $H(t)$ is modified back from $H(\tau_c^+)=H^{(Y)}T_h/T_c$ to $H(\tau_c+\tau_h)=H^{(X)}T_h/T_c$ in a time $\tau_h$. 
(iv) Finally a second quench is performed to restore $H^{(X)}T_h/T_c \rightarrow H^{(X)}$, closing the cycle. 

It is convenient to introduce the adimensional Hamiltonian $G(t):=\beta H(t)$, where $\beta=1/k_B T$ is the inverse temperature of the bath that the WS is coupled to (we shall set the Boltzmann constant $k_B$  equal to $k_B=1$). 
Note that the Carnot cycle becomes smooth with respect to $G(t)$, and in what follows, we take the driving protocol to be time-reversal symmetric, more precisely that $G(s\tau_c)=G(\tau_c+\tau_h(1-s))$ with $s\in [0,1]$. This property is always satisfied by optimal heat engines in the LD regime as long as  the two baths have the same spectral density \cite{slowdriving}, which is the subject of interest of this work. 
%we are interested in the  optimization of the cycle, and optimal engines have 
%of power as long as  the two baths have the same spectral density \cite{slowdriving}.
%(the asymmetric case is discussed in the Supp.Mat.%~\ref{SM-app:asym} \cite{SM}).   
By expressing $G(t)$ as 
\begin{align}
G(s\tau_c)=\sum_j \lambda_j(s) X_j, \hspace{8mm} s\in [0,1]
\label{time-symmetric}
\end{align}
 where $\lambda_j(s)$ are the control parameters and $X_j$ the conjugated forces, the cycle control can be characterised by $\tau_c$, $\tau_h$ and its shape $\vec{\lambda}(s)$ (notice that $\tau_c\neq\tau_h$ in general; the time-reversal symmetry is intended in the adimensional unit~$s$). We will not write explicitly the dependence on $s$, which will be clear from the context, but  will indicate with a dot the time derivative w.r.t. $s$, i.e. $\dot{G}\equiv\frac{\partial}{\partial s}G\equiv\tau_x\frac{d}{dt}G$, $x=(h,c)$.
 
We now assume the \emph{slow driving} or \emph{low dissipation} approximation, which is crucial in this work. 
 %The next assumption we do, which is crucial in this work, is the \emph{slow driving} or \emph{low dissipation} approximation. 
 That is, $\frac{d}{dt}{G}$ is finite but small, so that we can  expand the relevant quantities and keep  only leading terms (formally, the small parameter of the expansion is $\tau_{\rm eq}/\tau_x$%$\sim |\dot{\vec{\lambda}}|\tau_{\rm eq}$
, where $\tau_{\rm eq}$ is the relaxation time of the dynamics).  In this regime, the state of the WS is always close to thermal equilibrium, and the heat  exchanged during the isotherms (steps (i) and (iii)) can  be divided as 
\begin{equation}
\label{eq:lowdissipation}
Q_{x}=T_{x} \Delta S_{x}- W^{\rm diss}_{x}, \quad \quad x=(h,c)
\end{equation}
where $\Delta S_{x}$ is the reversible contribution obtained in the quasistatic limit $\tau_x\rightarrow \infty$, which is given by  $\Delta S \equiv \Delta S_h =-\Delta S_c=S(\omega(\tau_c))-S(\omega(0)) $ with $\omega(t)=e^{-G(t)}/\tr(e^{-G(t)})$ and where $S(\rho)$ is the Von Neumann entropy. The irreversible term $ W^{\rm diss}_{x}$ can be described in this regime by the so called \emph{thermodynamic length}~\cite{Schlogl1985,Salamon1983a,Nulton1,Crooks2007,Sivak2012a,Bonanca2014d,marti2018thermodynamiclength}
\begin{equation}
\label{eq:thermolength}
 W^{\rm diss}_{x}=\frac{T_{x}}{\tau_x}\int_0^{1} \sum_{ij} \dot{\lambda}_i m_{ij} \dot{\lambda}_j \ ds \ \equiv T_{x}\frac{\Sigma_x}{\tau_x} ,
\end{equation}
where  $m_{ij}$ is given, when the driven observables $\langle X_j \rangle$ relax to equilibrium with the same time-scale $\tau_{\rm eq}$, by~\cite{Schlogl1985,Salamon1983a,Nulton1,Crooks2007,Sivak2012a,Bonanca2014d,marti2018thermodynamiclength}:
\begin{align}
m_{ij}= \tau_{\rm eq}\frac{\partial^2 \ln \mathcal{Z}}{\partial \lambda_i \partial \lambda_j}.
\label{freeenergymetric}
\end{align}
where $ \mathcal{Z}= \tr{(e^{-G})}$ is the partition function. %For the sake of simplicity, here we have taken  $\tau_{\rm eq}$ to be constant along the trajectory, although this is not a necessary assumption. %Note that \eqref{freeenergymetric} is simply the response function (or susceptibility) of the WS when externally perturbed away from equilibrium, and indeed \eqref{freeenergymetric} gives the standard notion of thermodynamic length originally developed for macroscopic heat engines~\cite{Nulton1}. 
%Also, 
Given the time-reversal symmetry of the driving protocol, it follows that $\Sigma_h=\Sigma_c$ (\emph{symmetric low-dissipation regime} SLD). 
Importantly, while the results presented in the main  text are based upon the standard thermodynamic metric \eqref{freeenergymetric}, generalizations (including $\tau_{\rm eq}$ depending on the trajectory \cite{Sivak2012a,Bonanca2014d},  the possibility of having several relaxation time-scales, general Lindbladian dynamics~\cite{Sivak2012a,marti2018thermodynamiclength}, and protocols in which $\Sigma_h\neq \Sigma_c$)  are developed in the Supp. Mat.~\cite{SM}.

\emph{Optimisation of the cycle}. We now optimise the power (and efficiency) of the Carnot engine over  $\tau_c$, $\tau_h$ and its shape $\lambda_j(s)$, which are all the possible degrees of freedom. This enables us to obtain a fundamental upper bound on the power in the slow-driving regime, as well as the corresponding optimal control. 

The work extracted during a cycle is given by $W=Q_h+Q_c$, and the total time is $\tau=\tau_c+\tau_h$. The power hence reads $P=(Q_h+Q_c)/\tau$,
%\begin{align}
%P=\frac{Q_h+Q_c}{\tau},
%\end{align}
and the efficiency $\eta=(Q_h+Q_c)/Q_h$.
By substituting the expressions \eqref{eq:lowdissipation}-\eqref{eq:thermolength} and appropriately setting $\tau_c$ and $\tau_h$, one can maximize the power of the engine ($\partial P/ \partial \tau_j =0$) obtaining \cite{Schmiedl2007,esposito-EMP-bounds}
\begin{equation}
\label{eq:opt-pow_esposito}
P^{(\rm max)}=\frac{(\Delta S)^2}{4\Sigma}(\sqrt{T_h}-\sqrt{T_c})^2 
\end{equation}
and the corresponding efficiency at maximum power (EMP) is given by the Curzon-Ahlborn EMP, $\eta_{CA} =1-\sqrt{T_c/T_h}$ \cite{curzon-ahlborn}. 
%\begin{align}
%\label{etaCA}
%\eta_{CA} =1-\sqrt{\frac{T_c}{T_h}}. 
%\end{align}
In the most general case one might seek, in order to not sacrifice completely the efficiency optimization over the power, to maximize a hybrid figure of merit~\cite{Holubec2015,Holubec2016,Shiraishi2016,ma2018universal}.  The  maximum efficiency for any given power output of the engine
has been derived in \cite{ma2018universal} (see also \cite{Holubec2015}). Analogously, we can express the best power for a given efficiency, fixed to be a fraction of the maximum one: %$\eta=\gamma \eta_C, \hspace{8mm} \gamma \leq 1$,
\begin{align}
\eta=\gamma \eta_C, \hspace{8mm} \gamma \leq 1,
\label{eta}
\end{align} 
%$\eta=\gamma \eta_C$
 where $\eta_C =1-T_c/T_h$ is the Carnot efficiency. In the SLD regime this leads to a maximum power (cf. Supp.Mat.
 %~\ref{SM-app:asymptoticexp}
  \cite{SM}),
\begin{equation}
\label{res:max_pgamma}
P_\gamma^{(\rm max)}=\frac{(\Delta S) ^2}{4\Sigma}\frac{(T_h-T_c)^2\gamma(1-\gamma)}{\gamma T_c+(1-\gamma)T_h}\ ,
\end{equation}
obtained by setting
%\begin{align}
%&\tau_c=\frac{2\Sigma T_c}{\Delta S \left(T_h-T_c\right)(1-\gamma)}
%\nonumber\\
%&\tau_h=\tau_c\left( \frac{T_h}{T_c}(1-\gamma)+\gamma \right).
%\end{align}
$\tau_c=2\Sigma T_c / (\Delta S \left(T_h-T_c\right)(1-\gamma))$ and $\tau_h=\tau_c( T_hT_c^{-1}(1-\gamma)+\gamma )$.
Essentially, by tuning $\gamma$ in $\tau_c$ and $\tau_h $, one can  interpolate between a maximally powerful engine with power \eqref{eq:opt-pow_esposito} at efficiency $\eta_{CA}$, and a Carnot engine with maximal efficiency and null power. 

At this point, we note a crucial observation: after the optimisation of $P$  over $\tau_c$ and $\tau_h$, the remaining figure of merit is always ${(\Delta S) ^2}/{\Sigma}$, independently of the value of $\gamma$.  In fact, this is a property that can be argued to be general given any   figure of merit combining power and efficiency~\cite{SM}. %~\cite{Note2}. 
%\footnote{Given as objective any figure of merit $f(\eta,P)$ expressed in terms of the efficiency and the power, after optimization on $\tau_h,\tau_c$ the resulting value $\bar{f}:=f(\bar{\eta},\bar{P})$ (we indicate with $\bar{}$ quantities after time optimization) will be given, in the symmetric case, functions $\bar{\eta}=\bar{\eta}(T_h,T_C)$ and $\bar{P}=\hat{P}(T_h,T_c) (\Delta S) ^2 /\Sigma$ for some adimensional function $\bar{\eta}$ and homogeneous function $\hat{P}$ of the temperatures, by dimensional analysis. Moreover, any reasonable $f$ will be  monotonously increasing in both its parameters singularly (i.e. for a given efficiency we wish to enhance the power and vice-versa), therefore after speed optimization it will still be possible to improve the engine performance by increasing the ratio $(\Delta S)^2/\Sigma$.  The role of the above mentioned ratio as a characteristic scale defining the performance of both engines and refrigerators was already noticed in \cite{Hernndez2015} without further analysis.}. 
We now show how to maximize it by an opportune use of a Cauchy-Schwarz inequality. % by combining a geometrical approach to thermodynamics \cite{marti2018thermodynamiclength} with the Cauchy-Schwarz inequality. % (see Appendix XX).  %in the context of open quantum systems, although the the formalism can be extended to any system with dissipations that are linear in the speed of the control parameters.
%\emph{A geometric inequality to bound the power of low-dissipation Carnot engines}. 
First, using the formula for the derivative of an exponential~\cite{Hiai2014}: $\partial e^{- G}/\partial \lambda_j=-\int_0^1 e^{- G}X_j e^{-(1-s)G}$, we can express again $\Sigma$ in \eqref{eq:thermolength} in the more compact form
\begin{align}
\label{eq:sigma_std_metric}
\Sigma=\tau_{\rm eq}\int_0^1 ds \hspace{1mm}{\rm cov}_\omega (\dot{G},\dot{G})
\end{align}
where ${\rm cov}_\omega (A,B)$ is the Kubo-Mori-Boguliobov inner product:
%\begin{align}
${\rm cov}_\omega (A,B)=\int_0^1 ds \hspace{1mm} \tr(A\omega^{1-s}(B-\tr(\omega B))\omega^s)$. % (see e.g.~\cite{Dyson1978, Roepstorff}).
%\end{align}
 Next, we note that  the Von Neumann entropy $S=-\Tr[\omega \ln \omega]$ satisfies: $\dot{S}=-\tr(\dot{\omega}\ln \omega)=-{\rm cov}_\omega(\ln \omega,\dot{G})={\rm cov}_\omega(G,\dot{G})$ where we  used again the formula $\dot{\omega}=-\int_0^1 dx \hspace{1mm} \omega^{1-x}(\dot{G}-\tr(\omega \dot{G}))\omega^x$, as well as $\tr(\dot{\omega})=0$ and ${\rm cov}_\omega(A,\eye)=0$. Hence, we can write the total change in entropy $\Delta S$ as:
\begin{align}
\Delta S= -\int_0^1 ds \hspace{1mm} {\rm cov}_{\omega}(G,\dot{G}).
\end{align}
Crucially, both $\Delta S$ and $\Sigma$ can be expressed through an infinite-dimensional scalar product given by: $\langle A,B \rangle_\omega=\int_0^1 ds \hspace{1mm} {\rm cov}_{\omega}(A,B)$, that depends on the path $\{\lambda_j(s) \}$. % within the space of thermodynamic protocols where each infinite-dimensional vector is given by the shape of a cycle~$\{ \lambda_j(t)\}$.
 Using the Cauchy-Schwarz (C-S) inequality  $|\langle A,B \rangle|^2 \leq \langle A,A  \rangle \langle B,B \rangle$,  the ratio ${(\Delta S) ^2}/{\Sigma}$ can then be bounded as:
\begin{align}
\label{inequality}
\frac{(\Delta S) ^2}{\Sigma} \leq \frac{1}{\tau_{\rm eq}}\int_0^1 ds \hspace{1mm} {\rm cov}_{\omega}(G,G)\equiv \frac{1}{\tau_{\rm eq}}\int_0^1 ds \hspace{1mm} \mathcal{C}
\end{align}
where $\mathcal{C}$ is the \emph{heat capacity} of the WS,  
\begin{align}
\mathcal{C}= -\beta^2\frac{\partial^2 \ln\mathcal{Z}}{\partial \beta^2}
=\tr(\omega G^2) -\tr(\omega G)^2\,
\end{align}
i.e. the variance of the adimensional Hamiltonian $G$ for its thermal state. The C-S inequality is saturated for $\dot{G}\propto G $, which means that creating quantum coherence cannot favour the power output in the slow driving regime, in agreement with Refs.~\cite{Brandner2017,ma2018optim}. More importantly, it shows that optimal thermodynamic protocols take the simple form  $ G(t)=\lambda(t) G(0) $. 
%, which is often used in the literature (see e.g. \cite{}) without proving its optimality. Indeed, 
%As it often happens in thermodynamics, the simplest protocol turns out to  be the optimal one.  

We can further maximize \eqref{inequality} by  noting that $\int_0^1 ds \hspace{1mm} \mathcal{C}\leq \max_s \mathcal{C} $. To saturate this inequality in practice, 
one needs to consider  cycles where the modulation of $G$ is small
\begin{align}
&G(\tau_c)=G(0)(1+\epsilon) \hspace{5mm} \epsilon \ll 1,
\nonumber\\
&G(\tau_c+\tau_h)=G(0),
\label{OptSol}
\end{align} 
%\begin{align}
%&G(s \tau_c)= G(0)(1+\epsilon s\tau_c), \hspace{5mm} \epsilon \ll 1, \hspace{5mm} s \in [0,1]
%\nonumber\\
%&G(\tau_c+\tau_h(1-s))=G(s\tau_c)
%\label{OptSol}
%\end{align} 
around an optimal point $G(0)$ where $\mathcal{C}/\tau_{\rm eq}$ is maximised (as long as $\epsilon$ is small enough so that $G(s\tau_c)$ with $s\in[0,1]$ does not change substantially along the cycle, the precise form of $G(s\tau_c)$  is not important; see the SM for examples of explicit cycles).
In this case, in the limit $\epsilon\rightarrow 0$ the maximal power \eqref{res:max_pgamma}  for a given efficiency $\gamma \eta_C$ is  given by 
%\footnote{It might appear counterintuitive that one can obtain finite power with $\epsilon \rightarrow 0$. This can be understood by noting that both the work output $W$ and the time of the cycle $\tau$ are proportional to $\epsilon$.}
\begin{align}
\label{maxpower}
P^{\rm (max)}_\gamma= \frac{\mathcal{C}}{ 4\tau_{\rm eq}}\frac{(T_h-T_c)^2\gamma(1-\gamma)}{\gamma T_c+(1-\gamma)T_h}.
\end{align}
where $\mathcal{C}$ is the heat capacity at $G(0)$. We stress that \eqref{maxpower} has been obtained after maximising $P$ (for a fixed $\eta$) over all degrees of freedom: $\tau_c$, $\tau_h$ and the protocol $\{\lambda_j(s)\}$.  
% This is the main result of our work, whose generalizations are displayed in the SM, as it 
This result shows a fundamental link between maximal power of a finite-time Carnot cycle and the heat capacity of the WS.
%, as well as providing the corresponding optimal protocol.
 %simple expression for the maximal power of a finite-time Carnot cycle in the low dissipation regime. Indeed, 

 The simplicity of  \eqref{maxpower} can be contrasted  to exact optimisations of heat engines \cite{Schmiedl2007b,Armen2010,cavina_optimalcontrol,erdman2018maximum,menczel2019two}, where the full solution easily becomes too complex or not even computable with the size of the WS; and with other geometric optimisations which require solving geodesic equations to design optimal paths in the parameter space~\cite{Sivak2012a,Zulkowski2013,Zulkowski2015,Rotskoff2015,Rotskoff2017,marti2018thermodynamiclength,martimatteo_fluctuations}.
 % In the particular case where the WS is a single qubit, our result also agrees with  \cite{Esposito2010pre,abiuso_n-M}. 
 %we have reduced the initial optimisation over all cycles $\{ \lambda(t)\}$ to a maximisation of the heat capacity $\mathcal{C}$.  %as it provides a link between maximal power and the heat capacity of the WS. %This 
 In our approach,  from the point of view of optimization all that is left to do in~\eqref{maxpower} is to maximise~$\mathcal{C}$ over the control parameters~$\{\lambda_j\}$ to identify the optimal working point~$G(0)$ in~\eqref{OptSol}. % in order to find the optimal working point where to perform the infinitesimal cycles. 
In Fig.~\ref{fig:supra-extensive} explicit results are reported  for the value of maximum $\mathcal{C}$, for different paradigmatic levels of control on the same system of $N$ qubits.

Crucially, this approach can be generalized to any metric $m_{ij}$ in \eqref{eq:thermolength} describing dissipation: in the SM \cite{SM} we show that the optimal control problem is always reduced to infinitesimal cycles, and the optimal working point can be found by %an efficiently solvable
 a scalar maximisation problem. We work out as well examples of standard microscopical dynamics in open quantum systems: a qubit, a qutrit, or an   harmonic oscillator as a WS in contact with bosonic thermal baths with different spectral densities. %, or a  3-level system in contact with a bosonic bath.

It is important to point out that \eqref{maxpower}  should be understood as a theoretical ultimate upper bound on power,  obtained by taking $\epsilon \rightarrow 0$ in \eqref{OptSol}. In practice, any experimental or realistic protocol will   have finite $\epsilon$. In this case, as long as $\mathcal{C}$ is sufficiently smooth along the thermodynamic cycle \eqref{OptSol}, the power output $P$ (given by the integrated $\mathcal{C}$ in \eqref{inequality}) will not change considerably, so that realistic cycles will provide a similar $P$ than the theoretically maximal one.  In practice, keeping $\epsilon$ finite is also important to ensure  the consistency of the slow-driving approximation $\tau_{\rm eq}/\tau \ll 1$, given that for the optimized protocols that lead to the power \eqref{res:max_pgamma}, one has $\tau\propto \epsilon$, more precisely
\begin{equation}
\label{eq:consistency}
\tau_{\rm eq}/\tau\sim \epsilon^{-1}(T_h/T_c-1)(1-\gamma).
\end{equation} 
Note that this can always be guaranteed for  high efficiencies $\gamma\rightarrow 1$. From the same equation we note incidentally that engines
  whose maximum efficiency is constrained to be low ($T_c/T_h \lesssim 1$),
  i.e. arguably those engines that mostly need optimization in the high
  efficiency regime, show better convergence to the absolute
  bound.
%  \footnote{From the same equation we note incidentally that engines whose maximum efficiency is constrained to be low ($T_c/T_h \lesssim 1$), i.e. arguably those engines that mostly need optimization in the high efficiency regime, show better convergence to the absolute bound.} (see next section for examples). %We stress that \eqref{maxpower}  provides an upper bound on the power for any value of $\gamma$. 

%in practice as we show in the Supp.Mat.~\cite{SM} the optimal duration of the cycle scales as
%\begin{equation}
%\label{eq:consistency}
%\tau\sim \frac{\epsilon\tau_{\rm eq}}{(T_h/T_c-1)(1-\gamma)},
%\end{equation}
%hence for the slow-driving approximation to be consistent, the limit $\epsilon\rightarrow 0$ can be considered for asymptotically high efficiencies $\gamma\rightarrow 1$ \footnote{From the same equation we note incidentally that engines whose maximum efficiency is constrained to be low ($T_c/T_h \lesssim 1$), i.e. arguably those engines that mostly need optimization in the high efficiency regime, show better convergence to the absolute bound.}.}

\begin{figure}
\includegraphics[width=0.48\textwidth]{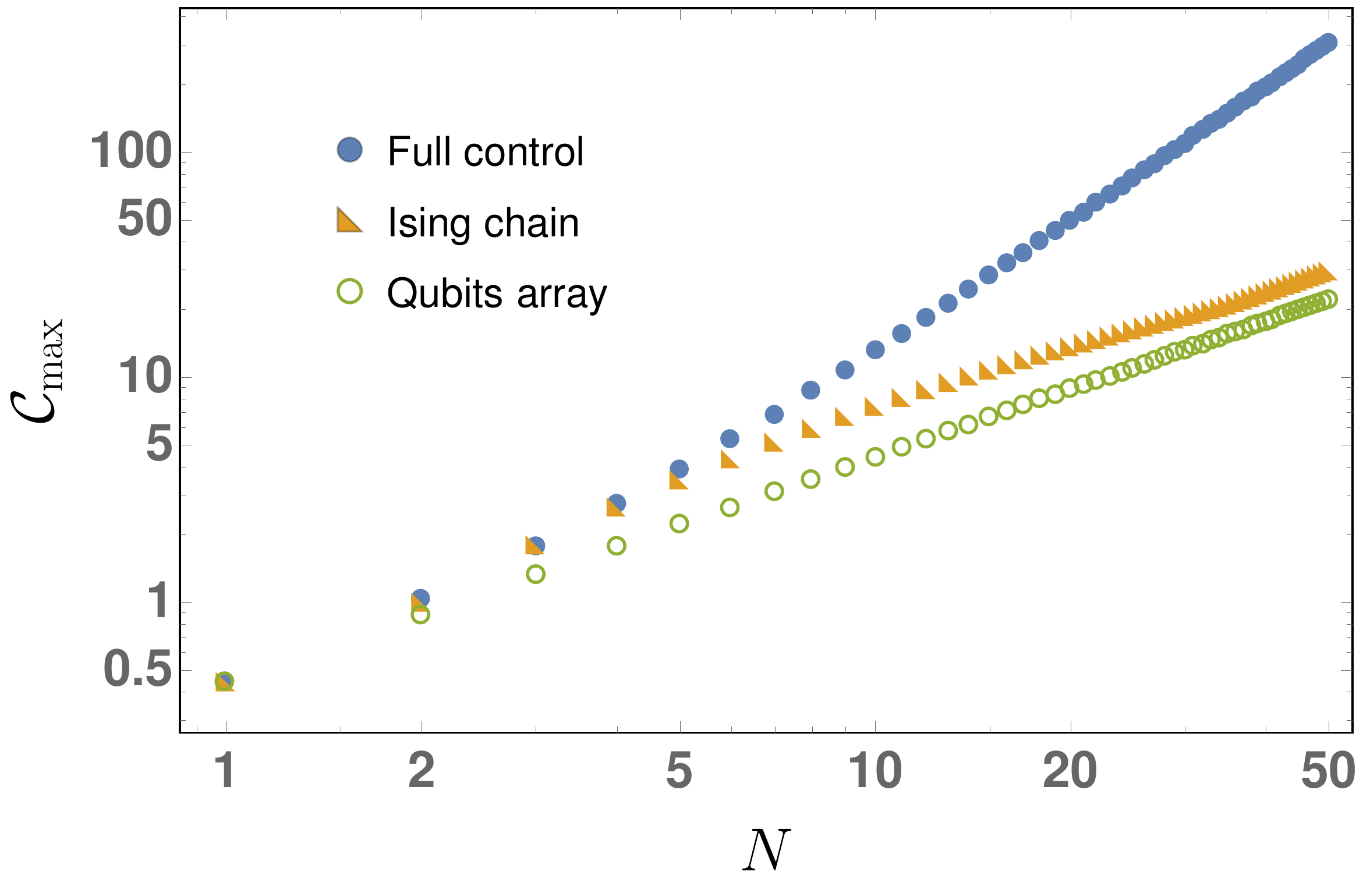}
\caption{Maximum  adimensional $\mathcal{C}$ for a thermal system of $N$ qubits with different degrees of control~\cite{SM}. \textbf{1)} $C_{\rm max} \simeq 0.44 N$ for  $N$ independent 2-level systems with gap control. \textbf{2)} For an Ising chain $H^{(N)}=-\lambda_1(t)\sum_{i=1}^N \sigma^z_i\sigma^z_{i+1}-\lambda_2(t)\sum_{i=1}^N\sigma^z_i$, we obtain $C_{\rm max} \simeq 0.59 N$. \textbf{3)}~Given complete control over the spectrum with $2^N$ levels,  $C_{\rm max}  \simeq N^2/4$. Details can be found in \cite{SM}. %When the Hamiltonian consists in $N$ non-interacting 2-level systems the optimal power is simply $N$ times the maximum power of a single qubit. In case of an Ising chain the set of possible Hamiltonians is enlarged (thus better power) but the asymptotic scaling remains linear. In case of full control of the $2^N$-dimensional Hamiltonian the maximum power scales as $N^2$
}
\label{fig:supra-extensive}
\end{figure}

\emph{Reaching Carnot efficiency at finite power.}
As an example of application of the previous results, we now use the designed optimal finite-time Carnot cycles (Eqs. \eqref{OptSol} where $G(0)$ will depend on each model of interest) to explore the possibility of reaching Carnot efficiency $\eta_C$ at finite power in the macroscopic limit. We follow the approach put forward in Refs. \cite{Finite_power_carnot_attainability,critical_engine}: considering a $N$-particle WS, we aim at approaching Carnot efficiency in the macroscopic limit $N\rightarrow \infty$ without giving up power per constituent. 

%Following \cite{Finite_power_carnot_attainability,critical_engine}, the crucial intuition is to  note that whenever the engine shows a supra-extensive scaling of the maximum power at finite efficiency, it is possible to sacrifice part of the power in order to obtain finite extensive power and Carnot efficiency in the limit of large size. 

To reach Carnot efficiency, we need $\gamma=1$ in \eqref{eta}, and hence we take $1-\gamma= N^{-\xi}, $
%\begin{align}
%&1-\gamma= N^{-\xi}, 
%\label{xi}
%\end{align}
where $\xi>0$ can be chosen at will. 
On the other hand, the maximal power $P_{\gamma}^{({\rm max})}$ in \eqref{maxpower} depends only on $\mathcal{C}$ and $\tau_{\rm eq}$; we then assume $\mathcal{C}= c_0 N^{1+a}$ and  $\tau_{\rm eq} = \tau_0 N^{b}$,
%\begin{align}
 %\mathcal{C}= c_0 N^{1+a},
%\quad
%\tau_{\rm eq} = \tau_0 N^{b},
%\end{align}
where the meaning of the different constants will later be described for each model of interest. Expanding the relevant quantities for $N\gg 1$, we obtain at leading order in $N$:
\begin{align}
\label{eq:asymptotic_expansionII}
&P_{\gamma}^{(\rm max)}=\frac{c_0(T_c-T_h)^2}{4\tau_0 T_c}N^{1+a-b-\xi}
\nonumber\\
&\tau_c= \frac{2\epsilon T_c}{T_h-T_c}\  \tau_0 N^{b+\xi}\ ,\hspace*{3.5mm}
\tau_h = \tau_c\ ,
\nonumber\\
&W=(T_h-T_c) c_0 \epsilon N^{1+a},
\hspace*{3.5mm}
\sigma_w^2=2 (T_h-T_c) \frac{W}{\epsilon}. 
\end{align}
where $\sigma_w^2=\langle w^2 \rangle -(\langle w \rangle)^2$ is the variance of the work distribution, which measures the work fluctuations per cycle of the engine (see Supp.Mat.
%~\ref{SM-app:asymptoticexp} 
\cite{SM} for details on the calculation). Let us now discuss two separate cases, inspired by   \cite{Finite_power_carnot_attainability} and \cite{critical_engine}, respectively.

{\bf (a) Control on the engine and the engine-bath interaction}. We first assume full control over the engine Hamiltonian with $2^N$ levels: that is, all levels can be modified at will by the experimentalist. While this is  extremelly challenging in practice, it is useful to obtain fundamental upper bounds on the maximal power. The optimal Hamiltonian maximising $\mathcal{C}$ then consists in a ground level
and a $2^{N}-1$ degenerate level (see \cite{Reeb2014,Correa-Mehboudi_optimalthermalprobes})  which, as shown in Fig.\ref{fig:supra-extensive}, leads to $\mathcal{C}\propto N^2$, i.e., $a=1$. 
This supralinear scaling is obtained in an increasingly smaller region of the parameter space, which requires taking $\epsilon \propto 1/N$ in \eqref{OptSol} (see details in Supp. Mat. \cite{SM}), and this constraints from Eq.~\eqref{eq:consistency} also $1-\gamma$ to scale accordingly, i.e. $\xi=1$. Furthermore, %as shown in Ref.\cite{Finite_power_carnot_attainability}, %the flat spectral configuration of the optimal Hamiltonian implies an exponential scaling of the thermalization timescale $\tau_{\rm eq}\sim e^N$  if the reservoirs are not fine-tuned to the engine. 
 it is possible to reach in realistic collisional scenarios $\tau_{\rm eq} \propto \sqrt{N}$ (i.e. $b=1/2$), or constant $\tau$ ($b=0$) if one is allowed to fine-tune the interaction between the WS and the baths \cite{Finite_power_carnot_attainability}%,Note4}.
  \footnote{We note that because of the degeneracies in the Hamiltonian of the WS the steady state might not be unique (as the population in each degenerate energy state might depend on the initial state). However, this is is irrelevant as we are considering a driving $\dot{G}\propto G$, where the power output depends only on the total population in each degenerate energy level. },
  % We obtain respectively $P_{1-\frac{1}{N}}^{(\rm max)} \propto N^{1/2}$ and  $P_{1-\frac{1}{N}}^{(\rm max)} \propto N$ in the two cases,
%In other words, one can obtain a convergence to Carnot with $\sim N^{-\alpha}$ with $\alpha<1/2$ while keeping the power output per particle constant.  
in agreement with Ref.~\cite{Finite_power_carnot_attainability}. In the Supp.Mat.~\cite{SM}, we solve exactly this proposal for a feasible driving
protocol close to the optimal one. 
%the maximum power scaling achievable at Carnot efficiency is found to be $O(N^{1/2})$ with $\xi=1$ 
%using an Otto-like cycle. 
%.  \textcolor{red}{Yet, the level of control in \cite{Finite_power_carnot_attainability}, a global unitary operation permuting the levels of the WS, is substantially higher  to the single-parameter driving $H=\lambda(t) H(0)$ considered here.}

 %substantial improvement of the result in Ref.\cite{Finite_power_carnot_attainability}, where the maximum power scaling achievable at Carnot efficiency is claimed to be $O(N^{1/2})$ under the same hypothesis. Note that the engine cycle used in  Ref.\cite{Finite_power_carnot_attainability} is reminiscent of an Otto cycle, whereas our design is based upon a (provably optimal) finite-time Carnot cycle. %in which the system only interacts with the thermal bath to thermalise, and it is brought far away from equilibrium in each adiabatic process. 

{\bf (b) Engine working on a phase transition point}. A promising platform to obtain supralinear scaling of power with realistic control is by choosing the engine to work in a phase transition point  of the many-body~WS.  For a finite
 WS operating close to the critical point, finite size scaling theory tells us that  $\mathcal{C}$ develops a peak of height $C \propto N^{1+\alpha/(\nu d)}$  and width $\delta \propto  N^{-1/(d\nu)}$, while $\tau_{\rm eq} \propto N^{z/d}$ (here  $\alpha$, $\nu$ and $z$ correspond to the specific heat, correlation length and dynamical critical exponents, while $d$ is the spatial dimension of the engine \cite{Fisher1972,Suzuki1977}). 
In order to exploit the critical scaling of the WS, we need to perform the cycle \eqref{OptSol} where $\mathcal{C}$ becomes peaked, and hence  $\epsilon \propto \delta \propto N^{-1/(d\nu)}$, which implies $\xi=1/(d\nu)$ from Eq.~\eqref{eq:consistency}.  Then, from \eqref{eq:asymptotic_expansionII} with $a=\alpha/(\nu d)$ and $b=z/d$,  supralinear scaling of $P_{\gamma}^{(\rm max)}$  is possible if %(details in the SM \cite{SM})
\begin{align}
\label{conditionphasetransition}
\alpha-z\nu -1 > 0\ .
\end{align}
This condition is the same found for the Otto cycle proposed in \cite{critical_engine}.
%{\color{red}A more detailed comparison between the Carnot cycle described here and the Otto cycle of \cite{critical_engine} is provided in the Supp. Material \cite{SM}.} %There, it is shown that the milder condition \eqref{conditionphasetransition} is possible 
%This difference  is a consequence 
Examples of physical systems where \eqref{conditionphasetransition} is satisfied are also provided in \cite{critical_engine}, particularly in the presence of critical speed-ups of thermalisation ($z<0$~\cite{Boukari1990,Grams2014,Tavora2014}). 

%, which hints on the realistic possibility of designing many-body heat engines working at Carnot efficiency with finite power. 

Besides efficiency and power, another crucial aspect of a heat engine is its reliability, i.e. the fluctuations in the output power. In fact, it has been recently pointed out in~\cite{critical_engine_fluctuations} that the Otto-cycle of~\cite{critical_engine} suffers from macroscopic fluctuations in the thermodynamic limit. For the Carnot-cycle considered here, from \eqref{eq:asymptotic_expansionII}  the relative work fluctuations read $f_w = \sigma_w /W =\sqrt{ 2(T_h- T_c) /\epsilon W}.$
%\begin{align}
%f_w = \frac{\sigma_w}{W}=\sqrt{ \frac{2(T_h- T_c) }{\epsilon W}}.
%\end{align}
First of all, in the case {\bf (a)} where $a=1$ and $\epsilon \propto N^{-1}$ in \eqref{eq:asymptotic_expansionII},  one has $f_w \sim \mathcal{O}(1)$ in the macroscopic limit. 
%: the fluctuations become comparable to the average extracted work per cycle.
 A similar situation takes place 
%one can easily achieve $f_w \rightarrow 0$ as $N\rightarrow \infty$ by taking $\epsilon\propto N^{-\zeta}$ with $0\leq \zeta<1$. Hence the fluctuations become negligible w.r.t. the extracted work for each cycle.
 %The case of 
 for the critical heat engine {\bf (b)}  as one simultaneously has  $\epsilon \propto N^{-1/(d\nu)}$  and $a=\alpha/(d\nu)$, and hence $f_w \propto N^{(-2+d\nu+\alpha)/(d\nu)}$. Using the relation $d \nu=2-\alpha$~\cite{huang2009introduction}, we hence obtain  $f_w \sim \mathcal{O}(1)$.
%~\ref{SM-app:asymptoticexp} ; and hence $f_w \sim \mathcal{O}(1)$. % (this is the same result found in \cite{critical_engine_fluctuations} for the Otto cycle of \cite{critical_engine}). 
Therefore, for both proposals $f_w \sim \mathcal{O}(1)$  in the thermodynamic limit, hence hindering their reliability, which is the same result found in \cite{critical_engine_fluctuations} for the Otto cycle. % \textcolor{blue}{Hence, for both the Carnot and the Otto cycle $f_w \sim \mathcal{O}(1)$  in the thermodynamic limit}. 
Despite $f_w \sim \mathcal{O}(1)$,  these fluctuations can be suitably avoided when the number $M$ of cycles is large %if $\tau_{\rm eq}$ is small compared to the total time $\tau_{\rm tot}$ of the output
(the argument below applies to the Otto and Carnot cycle).
 Given $M$ cycles, we  have that $f_w \propto 1/\sqrt{M}$ as the average work $W \propto M$ whereas the work fluctuations $\sigma_w \propto \sqrt{M}$  (think of a biased random walk). Therefore,  the ratio between the fluctuations per unit time and the power goes to zero as $M$ grows even when the fluctuations per single cycle are large. Since we have that  $M\propto\tau_{\rm tot}/\tau_{\rm eq}$, for a total time $\tau_{\rm tot}$ of observation, fluctuations can be suitably avoided  %, which for a fixed $\tau_{\rm tot}$, 
 e.g. for critical speed-ups where $\tau_{\rm eq}\propto N^b$ with $b<0$. 

%becomes possible if $\tau_{\rm eq}\propto N^b$ does not diverge with $N$, i.e. for $b=0$ or for $b<0$, corresponding to critical speed-ups of  thermalization.}  %if $\tau_{\rm ob}/\tau_{\rm eq}$ is large enough, e.g. in case of critical speedup of the thermalization $b<0$.

In actual implementations, the technical requirements to realise such optimised Carnot cycles are:  global control  of the WS, $H(t)=\lambda(t) H(0)$, and enough precision to engineer small cycles in the region where  $\mathcal{C}/\tau_{\rm eq}$ has supralinear scaling with $N$. Since the width $\epsilon$ of this region shrinks with $N$, in a realistic implementation the supralinear scaling will be eventually lost as the control precision is  limited~\footnote{Other possible implementation-dependent limitations may be the  cost of turning on/off the interaction with the baths \cite{PerarnauLlobet2016,Newman2017}, and a non-zero time for the quenches.}. %~\cite{Note5}.
  We remark that even when the experimental control may be
  limited, our considerations provide upper bounds on the maximal power of
  finite-time Carnot engines. %, as well as a general framework to efficiently

\emph{Conclusions.} We have characterized  the optimal cycle of a  finite-time Carnot engine  in the low-dissipation regime. The dissipation has been  characterized by the thermodynamic metric \eqref{freeenergymetric}, which is justified when the thermalization of the working substance (WS) is well described by a single time-scale~$\tau_{\rm eq}$. 
% (results can be extended to general metrics, see Supp.Mat. \cite{SM}). 
In this case, the optimal cycle  turns out to be remarkably simple: it consists of modulations in the form $\lambda(t) H(0)$, where $H(0)$ is the Hamiltonian of the WS. The power output is then proportional to the heat capacity $\mathcal{C}$ of the WS, linking the optimal performance to the nature of the WS: as an application we showed how the critical scaling of $\mathcal{C}$ can enable the design of optimal engines with extensive power reaching Carnot efficiency. These results have been generalised to general metrics in the Supp. Mat.~\cite{SM}, which we have used to derive the optimal cycle and corresponding power output of different WS (qubit, 3-level system, or harmonic oscillator) interacting with a bosonic thermal bath. Putting everything together, our reults provide a general framework to efficiently optimise the control of slowly driven  Carnot engines.

We hope this work stimulates further investigations in the  interplay between many-body physics and heat engines~\cite{Mascarenhas2014,critical_engine,critical_engine_fluctuations,Ma2017,Chand2018,
PerarnauLlobet2016,
Lekscha2018,Halpern2019,kloc2019collective,Jaramillo2016,Niedenzu2018}, as well as connections between performance, fluctuations, and degree of control \cite{Brandner2016}. 
In particular our results hint at the answer for two open problems: 1) small cycles are optimal for engine performance in all regimes \cite{cavina_optimalcontrol,abiuso_n-M,erdman2018maximum}, and 2) the performance of the proposals \cite{Finite_power_carnot_attainability,critical_engine} cannot be improved.

\emph{Acknowledgments}. 
We thank Harry J. D. Miller and Karen Hovhannisyan for insightful discussions. We also thank the anonymous referees for constructive criticisms and useful feedback.
P.A.  is  supported  by “la Caixa” Foundation (ID 100010434, fellowship code LCF/BQ/DI19/11730023), the  Spanish  MINECO  (QIBEQI  FIS2016-80773-P,  and Severo Ochoa SEV-2015-0522), Generalitat de Catalunya (SGR1381 and CERCA Programme), Fundacio Privada Cellex.

%%%%%%%%%%%%%%BIBLIOGRAPHY
\medskip
\bibliographystyle{apsrev4-1}

\bibliography{BIB.bib}

\newpage
\onecolumngrid
\pagebreak

\section*{Supplemental Material}
This Supplemental Material (SM) contains a generalisation of the results presented in the main text to open quantum systems, as well as technical details. We start in Sec. \ref{app:Carnot_cycle} with a review  the quantum Carnot cycle in the quasistatic limit. In Sec. \ref{app:thermolengths}  we introduce finite-time corrections for open quantum systems dynamics and show to characterise them geometrically. In Sec. \ref{SecApp:GenMainResult} we generalise the main result (the optimal cycle and corresponding maximal power) for general open quantum system dynamics by giving the solution as the maximisation of a scalar function. In Sec. \ref{sec:heatcap_models}, we present the maximisation of the heat capacity discussed in the main text. In Sec. \ref{sec:heatcap_models} we discuss the technical details of  the asymptotic expansions of power and efficiency presented in the main text. Finally, in Sec. \ref{SecAppExplicitSol} we solve analytically an illustrative   solvable case. 
To help the reader find information in this Supplemental Material we introduced Table \ref{tb:1}. 

\begin{table}[h]
\begin{tabular}{l|l}
Section I & Description of a quasistatic, reversible Carnot cycle for a general quantum system.\\ \hline
Section II & Introduction of finite-time Carnot cycle and thermodynamic length.\\
IIA & Thermodynamic metric for exponential relaxations of the observables.\\
IIB & Thermodynamic metric from microscopical models.\\ 
IIB1 & Example: Qubit with bosonic baths.\\ 
IIB2 & Example: Harmonic oscillator with bosonic baths.\\ 
IIB3 & Example: 3-level system with detailed balance evolution.\\ \hline
Section III & Generalizations of cycle optimization to general metrics and protocols.\\
IIIA & Generalization for asymmetric dissipation and symmetric protocols.\\
IIIA1 & Optimization for systems with point-dependent thermalization timescale.\\
IIIA2 & Optimization for general metrics.\\
IIIA3 & Solution of the optimization for the models presented in Section IIB1-2-3.\\
IIIB & Bounds on completely asymmetric protocols.\\ \hline
Section IV & Maximization of the heat capacity for systems with different degree of control.\\ \hline
Section V & Time-tuning optimization of a low-dissipation Carnot engine.\\
VA & Critical scalings of power, efficiency and fluctuations of optimal protocols.\\
VB & Comparison with  Otto cycle of Ref.\cite{critical_engine_fluctuations}\\ \hline
Section VI & Explicit analytical control protocol for a $D-1$ degenerate model.
\end{tabular}
\caption{Guide table for the Supplemental Material.}
\label{tb:1}
\end{table}

\section{Quasistatic quantum Carnot cycle}
\label{app:Carnot_cycle}
For completeness, in this first section we review  the quantum Carnot cycle in the quasistatic limit (i.e. without including finite-time corrections).

%\subsection{Quasistatic quantum Carnot cycle}
The internal energy of a system with Hamiltonian $H$ in the state $\rho$ is defined as 
\begin{equation}
U=\Tr[\rho H]\ .
\end{equation}
Considering the variation $d U$, it is possible to identify \cite{alicki1979,anders-giova_discrete,kieuWork,anders_thermoreview} the work and heat contributions 
\begin{align}
d W=\Tr[\rho\  d H]\ , \\ 
\label{def:heat}
d Q=\Tr[d\rho \ H]\ .
\end{align}
To simulate sensible restraints on the system, external control is assumed on dynamical parameters of the local Hamiltonian of the system $H_t=H(\vec{l}(t))$. When in contact with a reservoir at temperature $T=1/\beta$ (we use units in which the Boltzmann's constant is $k_B=1$), such a system relaxes to the the Gibbs state 
\begin{equation}
\label{def:gibbs_state}
\omega_\beta(\vec{l})=e^{-\beta H}/\mathcal{Z}_\beta(H) 
\end{equation} 
(here $\mathcal{Z}_\beta(H)=\Tr[e^{-\beta H}]$ is the partition function of the system in the canonical ensemble).

\begin{figure}
\includegraphics[scale=0.6]{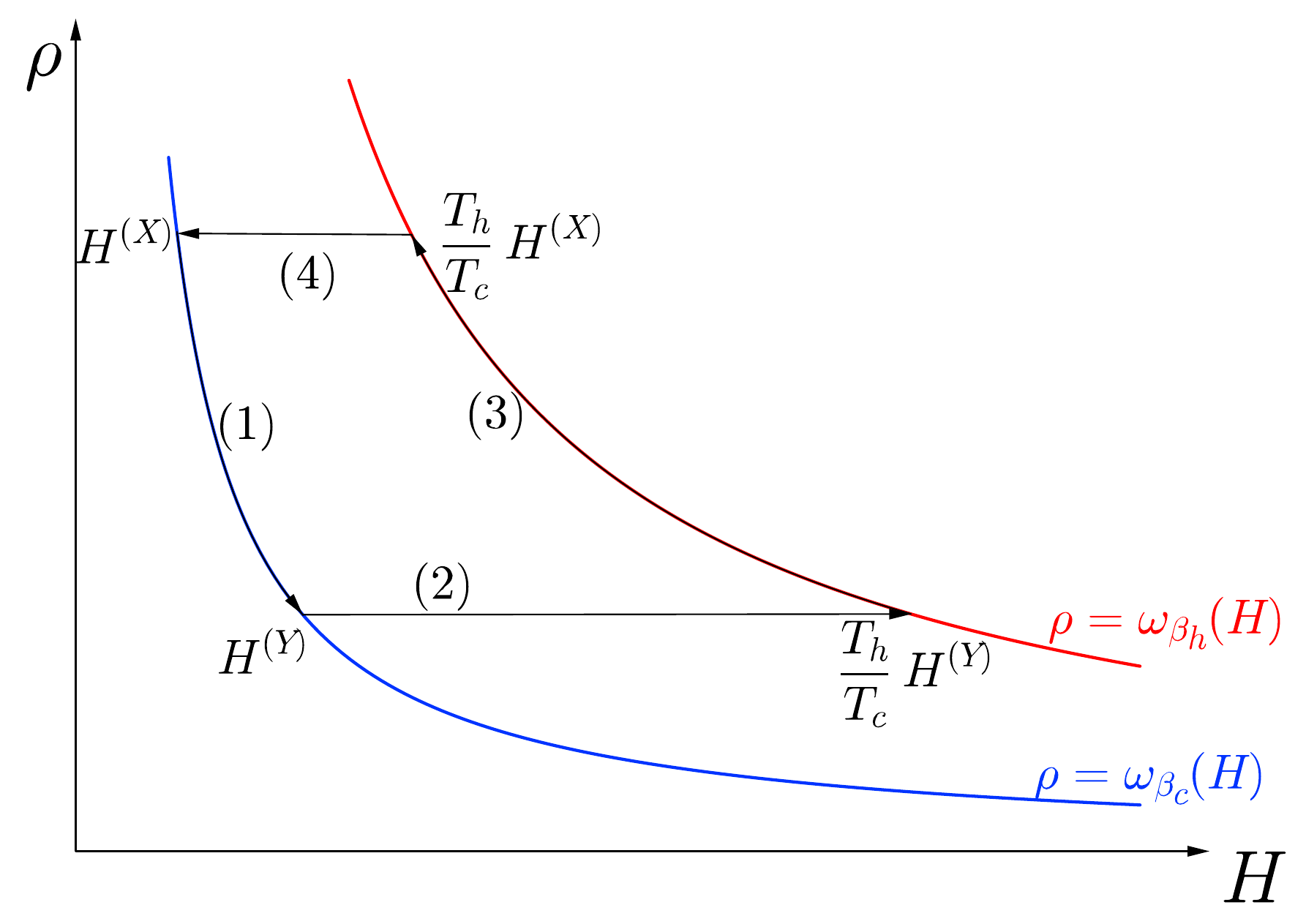}
\caption{Pictorial, bidimensional representation of a Carnot cycle, in the "phase space" defined by its state and Hamiltonian: two isothermal strokes, where the Hamiltonian is modified and the system approximately follows the Gibbs state, alternated with two rapid quenches where the system keeps the same state.}
\label{fig:Carnot}
\end{figure}

A \emph{Carnot Cycle} \cite{anders_thermoreview,abiuso_n-M,cavina_optimalcontrol,slowdriving} is identified with a 4 steps process, that is two isothermal strokes alternated with two isoentropic (adiabatic) strokes (cf. Fig.~\ref{fig:Carnot}).
Consider a system with a controlled Hamiltonian $H_t$
which can be coupled independently to two reservoirs with temperature $T_h>T_c$. In the ideal \textit{quasistatic limit} the operations are performed slowly enough to allow the system to be in thermal equilibrium $\rho(t)\equiv\omega_\beta(\vec{l}(t))$ at every instant.
The 4 steps are:
\begin{itemize}
\item [1)] while being coupled to the cold reservoir, the Hamiltonian is modified continuously from $H^{(X)}$ to $H^{(Y)}$ such that  $\Tr[\dot{\omega}_\beta H_t]$ is negative, in order for heat to be released to the cold source.
\item [2)] with the system isolated from the reservoirs, a quench is performed taking $H^{(Y)}\rightarrow H^{(Y)}\frac{T_h}{T_c}$.
\item [3)] while being coupled to the hot reservoir, the Hamiltonian is modified continuously from $H^{(Y)}\frac{T_h}{T_c}$ to $H^{(X)}\frac{T_h}{T_c}$.
\item [4)] again isolating the system a quench is performed to restore $H^{(X)}\frac{T_h}{T_c}\rightarrow H^{(X)}\frac{T_h}{T_c}\frac{T_c}{T_h}=H^{(X)}$.
\end{itemize}
Note the factors $\frac{T_h}{T_c}$ and $\frac{T_c}{T_h}$ are chosen in order for the state to be continuous during the quenches. In fact the thermal state uniquely depends on $\beta H_T$ (cf.~\eqref{def:gibbs_state}); i.e. for example during the quench 2) the relation $\beta_c H^{(Y)}=\beta_h H^{(Y)}\frac{T_h}{T_c}$ guarantees $\omega_{\beta_c}(H^{(Y)})=\omega_{\beta_h}(H^{(Y)}\frac{T_h}{T_c})$. We shall thus define then the adimensional Hamiltonian at temperature $1/\beta$ (note that the temperature takes only two values, that depend on the respective isotherms)
\begin{equation}
\label{def:adimensional_hamiltonian}
G_{t'}:=\beta H_{t'\tau}
\end{equation}
so that the thermal state is $\omega_t'=e^{-G_{t'}}/\Tr[e^{-G_{t'}}]$ on both the cold and hot isotherm. The time reparametrization is conceived in order to isolate the shape of the control $G_{t'}$, with $0\leq t' \leq 1$. Note that $G$ (and hence $\omega$) is continuous also on the quenches; that is, we can consider the cold isotherm consisting in a transformation $(\omega^{(X)},G^{(X)})\rightarrow (\omega^{(Y)},G^{(Y)})$ with duration $\tau_c$ and the hot isotherm the opposite $(\omega^{(Y)},G^{(Y)})\rightarrow (\omega^{(X)},G^{(X)})$, with duration $\tau_h$. In a Carnot cycle heat is exchanged only during the 1), 3) steps (absorbed from the hot source, released to the cold one), hence we can compute the efficiency using the observation that over a cycle $\Delta Q+\Delta W=0$ and the heat is absorbed from the hot bath $Q^{(0)}_{abs}=Q_h^{(0)}$ (we use the superscript$^{(0)}$ to indicate quantities in the quasistatic regime)
\begin{equation}
\eta_{Carnot}=\frac{-\Delta W^{(0)}}{Q^{(0)}_{abs}}=1+\frac{Q_c^{(0)}}{Q_h^{(0)}}=1-\frac{T_c}{T_h}\ ,
\end{equation}
where we have used in the last step that in the quasistatic limit the above mentioned considerations together with Eq.~\eqref{def:heat} give
\begin{equation}
\frac{Q_c^{(0)}}{T_c}=\int_X^Y\Tr[d\omega\ G]=-\int_Y^X\Tr[d\omega\ G]=-\frac{Q_h^{(0)}}{T_h}\ .
\end{equation}

\section{Finite-time Carnot cycle and thermodynamic length}
\label{app:thermolengths}

We now consider finite-time corrections  on the above quasistatic Carnot cycle through a quantum open system approach, where the dissipation in linear response can be described in a geometric form by following \cite{marti2018thermodynamiclength}. We consider an isothermal process where the Hamiltonian $H(t)$ of the WS is driven for $t\in [0,\tau]$ in contact with a thermal bath at temperature $T=1/\beta$. In order to characterise the process beyond the quasistatic limit, we need to assume some structure on the thermalization processes of the working substance (WS) induced by the reservoirs.  In a rather generic scenario we consider that the relaxation of the WS can be described by a master equation with the thermal state as a unique fixed point:
\begin{equation}
\label{eq:general_QTmachineEq}
\frac{d}{dt}{\rho}=\LL[\rho] \quad \text{s.t.} \quad \LL[\omega_\beta(H(t))]=0.
\end{equation} 
where we note that $\LL \equiv \LL(t)$ is also time-dependent. 
Following \cite{slowdriving}, the solution of this master equation  can be found perturbatively around $1/\tau$. Using the renormalised time $s \in [0,1]$ with  the convection $H_s \equiv H(s\tau)$ and $\rho_s \equiv \rho(s\tau)$, the solution is given by (indicating with a dot $\dot{}$ the time derivative w.r.t. $s$):
\begin{align}
\rho_{s}= \omega_{s} + \frac{1}{\tau} \LL_{s}^{-1}[\dot{\omega}_{s}]+...
\end{align}
with $\omega_s = e^{-\beta H_s}/\tr(e^{-\beta H_s})$, and 
% by expressing the heat released/absorbed by the baths as
%\begin{align}
%\label{lowdisapp}
%&Q_h=T_h \Delta S - \frac{T_h\Sigma_h}{\tau_h}
%\nonumber\\
%&Q_c=-T_c \Delta S - \frac{T_c\Sigma_c}{\tau_c}
%\end{align}
where $\LL_{s}^{-1}$ is the inverse of $\LL_{s}$ within the traceless subspace of density matrices, the so called Drazin inverse (see e.g. \cite{crooks2018drazin,marti2018thermodynamiclength,abiuso_n-M,slowdriving}). Plugging this expression into $Q= \int_0^1 ds \hspace{1mm} \Tr(\dot{\rho}H) $ and using integration by parts we obtain:
\begin{align}
Q=T \Delta S - \frac{T}{\tau} \Sigma
\end{align} 
with the first order correction to the quasistatic limit given by:
\begin{align}
\Sigma=\beta \int_0^1 ds \hspace{1mm} \Tr( \mathcal{L}_{s}^{-1}[\dot{\omega}_{s}]\dot{H}_s) 
\end{align}
In terms of the adimensional Hamiltonian  $G_s \equiv \beta H_s$, and using the formula for the derivative of an exponential,
\begin{align}
\label{def:J}
\dot{\omega} = -\int_0^1 dx \hspace{1mm} \omega^{1-x}(\dot{G}-\tr(\omega \dot{G}))\omega^x\equiv \mathfrak{J}_\omega[\dot{G}],
\end{align} 
we can write the variation of entropy as
\begin{equation}
\Delta S=-\int_0^1 ds \Tr[\mathfrak{J}_{\omega_s}[\dot{G}_s]\ln\omega_s]=\int_0^1 ds \Tr[G_s\mathfrak{J}_{\omega_s}[\dot{G}_s]]\ ,
\end{equation}
while $\Sigma$ can be reexpressed in the convenient quadratic form: 
\begin{equation}
\label{sigmageneral}
\Sigma=-\int_0^1 \D s \Tr[\dot{G}_{s}\mathcal{L}_{s}^{-1}[\mathfrak{J}_{\omega_{s}}[\dot{G}_{s}]]] \ .
\end{equation} 
Now, expanding $G_s$ as  
\begin{align}
\label{ExpG}
G_{s} = \sum_j \lambda_j(s) X_j \ ,
\end{align}
we can conveniently write $\Delta S$ and $\Sigma$ as
\begin{align}
\Delta S=\sum_i\int_0^1 \D s\  s_i\dot{\lambda}_i \ ,\qquad\Sigma= \sum_{ij} \int_0^1 {\rm d}s \hspace{1mm}\dot{\lambda}_i m_{ij} \dot{\lambda}_j
\label{SigmaslowdrivingIII}
\end{align}
with
\begin{align}
s_i=\Tr[G_s\mathfrak{J}_{\omega_s}[X_i]] \qquad \text{and} \qquad m_{ij}= -\frac{1}{2} \left(\Tr(X_i \mathcal{L}_{s}^{-1} \mathfrak{J}_{\omega_{s}}[ X_j]+ X_j \mathcal{L}^{-1}_{s} \mathfrak{J}_{\omega_{s}}[ X_i]) \right).
\label{mgeneral}
\end{align}
The matrix $m_{ij}$ is symmetric, positive-definite due to the second law $d\Sigma \geq 0$, and it depends
smoothly on the base point $\omega$; hence it defines a metric.

Finally, we note that  the linear expansion \eqref{SigmaslowdrivingIII}  can also be obtained in exact treatments of the system-bath Hamiltonian dynamics through  linear-response theory~\cite{Campisi2012geometric,Bonanca2014d,acconcia2015shortcuts,Ludovico2016adiabatic}. As a rule, the expansion \eqref{SigmaslowdrivingIII} can be argued to be general for any system with dissipations that are linear (at the lowest order) in the speed of the driving. Suppose indeed that the dissipation along an infinitesimal segment of the trajectory depends only on the local point and the local driving. As a consequence it must be in the form $d\Sigma=d\lambda_i f_i(\lambda,\dot{\lambda},\ddot{\lambda},...)$, but the $1/\tau$ scaling implies that the first derivative terms enter linearly in the product while higher orders are suppressed, which implies $d\Sigma=d\lambda_i m_{ij}(\lambda)\dot{\lambda}_j$, which is equivalent to ~\eqref{SigmaslowdrivingIII}.
%Note also that $m_{ij}$ is a positive-definite matrix due to the second law, $d\Sigma >0$. Later in the text we will consider specific cases of $m_{ij}$, yet at the moment we keep it general to ensure a wide applicability of our result.

\subsection{Thermodynamic metric for single or multiple time scales}
\label{sec:STD_thermolength}
We first show how to obtain the Kubo-Mori-Boguliobov metric used in the main text (Eqs. (4) and (8)). This can be easily done by taking an exponential relaxation of $\rho$ to equilibrium with a single time-scale, as described by the Lindbladian: 
\begin{align}
\LL_s[\rho_s]= \tau_{\rm eq}^{-1} (\omega_s-\rho_s)
\label{trivialLind}
\end{align}
which has the Drazin inverse $\LL_{s}^{-1}[.]=\tau_{\rm eq}(\omega_s\Tr(.)-\mathbb{I})$. In this case, using that $\Tr (\mathfrak{J}_\omega[\dot{G}])=0$ one finds that $\Sigma$ in \eqref{sigmageneral} is given by:
\begin{equation}
\label{sigmapart}
\Sigma=\tau_{\rm eq}\int_0^1 \D s \Tr[\dot{G}_{s}[\mathfrak{J}_{\omega_{s}}[\dot{G}_{s}]]] \ .
\end{equation} 
that is Eq.~(8) from the main text, as the generalised covariance is given by 
${\rm cov}(A,B)= \Tr[A[\mathfrak{J}_{\omega_{s}}[B]]]$. We hence see that the standard thermodynamic length (Eq. (4) in the main text) can be obtained by an heuristic model of thermalisation with a single time-scale. % (originally, it was derived by considering perturbations on the thermal state \REF). 

It is important to keep in mind that these considerations are only relevant close to equilibrium, where the metric (4) in the main text becomes a good thermodynamic description. Furthermore, although we have assumed that the whole state $\rho$ converges to equilibrium as in \eqref{trivialLind}, strictly speaking it is only necessary that the driven observables (the $X_j$ in \eqref{ExpG}) converge to equilibrium with a single time-scale, i.e.,  $\langle \dot{X}_i \rangle_{\rho} = \tau_{\rm eq}^{-1} (\langle X_i \rangle_{\omega} - \langle X_i \rangle_{\rho})$,  with $\langle X \rangle_\rho = \tr(X\rho)$. This is especially relevant in complex systems (e.g. many-body systems), where the full dissipative dynamics can be extremely complex but the equilibration of some macroscopic observables can be well described by an exponential relaxation with a suitable time-scale. In this sense, it is also worth pointing out that if each generalised observable $X_i$ decays with a different time-scale $\tau_i$ 
\begin{equation}
\label{eq:multiple_relax}
\langle \dot{X}_i \rangle_{\rho} = \tau_{i}^{-1} (\langle X_i \rangle_{\omega} - \langle X_i \rangle_{\rho}) \ ,
\end{equation}
then the metric Eq. (4) of the main text can be easily generalised as~\cite{marti2018thermodynamiclength}: 
\begin{align}
m_{ij}= \frac{\tau_i+\tau_j}{2}\frac{\partial^2}{\partial \lambda_i \partial \lambda_j} \ln \mathcal{Z},
\label{metricmij}
\end{align}
  where we have absorbed the dependence on $\tau_i$ in the metric and where $\tau_i$ can in principle depend on the point of the trajectory.   
  
  As a final remark, we note that while we have derived the metrics \eqref{sigmapart} and \eqref{metricmij} with an heuristic approach based on exponential relaxation near equilibrium, one can derive them from a microscopic derivation based upon linear-response  \cite{Sivak2012a,Bonanca2014d}, in which case $\tau_{i}$ will in general depend on the point of the trajectory, i.e., $G$. This case will be treated in Sec. \ref{app:C-S}.

\subsection{Thermodynamic metric on standard microscopical models}
In this section, we apply our general considerations to derive thermodynamic metrics for systems described  by quantum master equations. % that can be derived from standard physical models.
 Before, let us discuss here some generic properties.
Following  \cite{slowdriving,breuer-petruccione}, we have in standard scenarios where a quantum system is coupled to a bosonic bath:
\begin{align}
\LL_s[\rho_s]= \sum_{\nu>0}\gamma_0 \nu^{\alpha} \left(  (N(\beta\nu)+1) \mathcal{D}_{A_\nu}[\rho_s] +   N(\beta\nu)\mathcal{D}_{A_\nu^{\dagger}}[\rho_s] \right)
\label{Lindmic}
\end{align}
where $\gamma_0 \nu^{\alpha}$ is the spectral density of the bath ($\alpha$ defines the Ohmicity), 
\begin{align}
 \mathcal{D}_{X}[\rho]= X \rho X^{\dagger}-\frac{1}{2}(X^{\dagger}X \rho+ \rho X^{\dagger}X)
\end{align}
and
\begin{align}
N(\beta \nu)= \frac{1}{e^{\beta \nu}-1}
\label{Nw}
\end{align}
is the Bose-Einstein distribution (one can consider ferimonic baths by replacing $N(\beta\nu)$ by the Fermi distribution). 
The general form \eqref{Lindmic} of  $\mathcal{L}$ depends on $H_s$, $\beta$ and the spectral density $J(\nu)=\gamma_0 \nu^{\alpha}$.
\label{sec:spectral_dens}
Consider now a Carnot cycle with two baths at temperature $\beta_c$ and $\beta_h$, and the action of each bath being described by a $\mathcal{L}(H_S, \beta, J(\nu))$. Assuming that the spectral density of both baths is the same, then both Lindbladians are related by the transformation  $\beta \rightarrow \lambda^{-1}
 \beta$ and $H_S \rightarrow \lambda H_S$, with $\beta \equiv \beta_c$ and $\lambda \equiv \beta_c/\beta_h$.
 In terms of the Lindbladian, note that~\cite{slowdriving}:
 \begin{align}
 \label{relationLind}
 \mathcal{L}(\lambda H_S , \lambda^{-1} \beta) =  \lambda^{\alpha}\mathcal{L}( H_S ,  \beta),
 \end{align}
 which shows how the generators of the dynamics are related between the cold and hot isotherm, i.e. 
 \begin{align}
 \mathcal{L}(({\beta_c}/{\beta_h}) H_S , \beta_h) =  \bigg(\frac{\beta_c}{\beta_h}\bigg)^{\alpha}\mathcal{L}( H_S ,  \beta_c) \ .
 \end{align}
 After noticing that the dissipation is related to $\mathcal{L}$ through Eq.~\eqref{sigmageneral}, we can write a simple proportionality relation for the metric describing the dissipation on the hot and cold isotherm,
\begin{equation}
m^{(h)}_{ij}=\bigg(\frac{T_c}{T_h}\bigg)^{\alpha}m^{(c)}_{ij} \ ,
\end{equation} 
which implies, for time-reversal symmetric protocols described in the main text,
 \begin{equation}
 \Sigma_h=\bigg(\frac{T_c}{T_h}\bigg)^{\alpha}\Sigma_c \ .
 \end{equation}
 While from this relation we see that the considerations of the main text (for symmetric protocols $\Sigma_c=\Sigma_h$) in principle only apply for flat spectral densities $\alpha=0$, we will show in Sec. \ref{sec:proportional_diss} that the same figure of merit $\Delta S^2 /\Sigma_c$ (or equivalently $\Delta S^2 /\Sigma_h$) needs to be maximised to obtain maximal power. 
 
An important comment is now in order. Whereas the dissipator \eqref{Lindmic} is usually derived for time-independent Hamiltonians \cite{breuer-petruccione}, here we are interested in slowly driven Hamiltonians. Nevertheless, the same form for the dissipator is justified  as long as  the bath dynamics are fast compared to the driving rate of the system Hamiltonian, which leads to the well-known adiabatic master equation \cite{Albash2012,Dann,Yamaguchi2017}. 
 %Indeed, identifying $\beta \equiv \beta_c$ and $\lambda \equiv \beta_c/\beta_h$, we recover the structure  of the Carnot cycle. 
%For future convenience, we note that $\LL_s[\rho_s]$ depends  on the temperature through  \eqref{Nw} and the following limits:
%\begin{itemize}
%\item For high temperatures, $\beta \hbar w \ll 1$, then $N(w) \approx (\beta \hbar w)^{-1}$. Note that in this case the general metric \eqref{mgeneral} depends on the temperature through an overall factor $\beta$. 
%\item For low temperatures, $\beta \hbar w \gg 1$, then $N(w) \approx 0$, and the metric \eqref{mgeneral} does not depend on $\beta$. 
%\end{itemize}
\subsubsection{Two-level system with a bosonic bath}
\label{subsec:qbit_mij}
The well-known optical master equation \cite{breuer-petruccione} can describe a two-level system with Hamiltonian $G\equiv\beta H=\frac{w}{2}\sigma^z$ relaxing in a bosonic thermal bath at temperature $1/\beta$, and from \eqref{Lindmic} and  working in the interaction picture it takes the form 
\begin{equation}
\label{eq:optical-MME-qbit}
\dot{\rho}=\Gamma(1+N(w))\big(\sigma^-\rho\sigma^+-\frac{1}{2}\{\sigma^+\sigma^-,\rho\}\big)+\Gamma N(w)\big(\sigma^+\rho\sigma^--\frac{1}{2}\{\sigma^-\sigma^+,\rho\}\big)\ ,
\end{equation}
where  the rate $\Gamma\propto w^\alpha$ depends on the ohmicity of the bath (for a standard atomic-optical field interaction $\alpha=3$ \cite{breuer-petruccione}) and $\sigma^\pm=(\sigma^x\pm i\sigma^y)/2$. It is easy to translate this equation on the single Bloch-vector components
\begin{subequations}
\begin{align}
\frac{d}{dt}\langle{\sigma}^x\rangle&=- \frac{\Gamma(2N+1)}{2} \langle\sigma^x\rangle\ ,\\
\frac{d}{dt}\langle{\sigma}^y\rangle&=- \frac{\Gamma(2N+1)}{2} \langle\sigma^y\rangle\ ,\\
\frac{d}{dt}\langle{\sigma}^z\rangle&=-\Gamma(2N+1) \langle\sigma^z\rangle- \Gamma \ .
\end{align}
\end{subequations}
This equations being in the form \eqref{eq:multiple_relax}, imply that the thermodynamic metric is indeed in the form \eqref{metricmij}. Moreover the covariances of different cartesian components decouple as ${\rm cov}(\sigma^i,\sigma^j)\propto\delta_{ij}$ (with $i,j,=x,y,z$) hence implying 
\begin{equation}
\label{eq:optical_qubit_mij}
m_{ij}=
\frac{1}{\Gamma(2N(w)+1)}
\begin{pmatrix}
2{\rm cov}(\sigma^x,\sigma^x) & 0 & 0 \\
0 & 2{\rm cov}(\sigma^y,\sigma^y) & 0 \\
0 & 0 & {\rm cov}(\sigma^z,\sigma^z)
\end{pmatrix}
\end{equation}
Consider now the external control on the qubit, i.e. ($\hat{n}(t)$ is a unit vector)
\begin{equation}
G(t)\equiv\beta H(t)\frac{w(t)}{2}\hat{n}(t)\cdot\vec{\sigma}.
\end{equation}
Assuming that the driving is much slower than the internal dynamics of the bath (i.e. the adiabatic master equation \cite{Albash2012,Dann,Yamaguchi2017}), the same master equation \eqref{eq:optical-MME-qbit}  and metric \eqref{eq:optical_qubit_mij} instantaneously holds in the rotated basis where $\sigma^z(t)=\hat{n}(t)\cdot\vec{\sigma}$ and with $\omega\rightarrow \omega(t)$ becoming time dependent. 
The same metric is expressed in polar coordinates in \cite{marti2018thermodynamiclength}.  
%If in the same model we assume the system to be in contact with a fermionic field at thermal equilibrium instead of a bosonic one, the fermionic commutation relations make the structure of Eq. (\ref{eq:optical-MME-qbit}) change in \REF
%\begin{equation}
%\label{eq:optical-fermionic_eq.}
%\dot{\rho}=\Gamma(1-N)\big(\sigma_-\rho\sigma_+-\frac{1}{2}\{\sigma_+\sigma_-,\rho\}\big)+\Gamma N\big(\sigma_+\rho\sigma_--\frac{1}{2}\{\sigma_-\sigma_+,\rho\}\big)
%\end{equation}
%with $N(w)=\frac{1}{e^{\beta w}+1}$  in  this case.
%This master equation can be used to derscibe thermalising effects involving fermionic systems like quantum dots \cite{harbolaMME}. \\
%\begin{subequations}
%\begin{align}
%\dot{a}_1&=- \frac{\Gamma}{2} a_1\ ,\\
%\dot{a}_2&=- \frac{\Gamma}{2} a_2\ ,\\
%\dot{a}_3&=-\Gamma a_3- (2N-1)\Gamma \ .
%\end{align}
%\end{subequations}
\subsubsection{Harmonic oscillator with a bosonic bath}
\label{simple_harmosc_metric}
As in the previous example, the optical master equation can be written similarly for an harmonic oscillator with gap control $\beta H(t)=w(t) a^{\dag}a$ (again working on the interaction picture and assuming an adiabatic master equation)
\begin{equation}
\label{eq:optical-MME-harmonic}
\dot{\rho}=\Gamma(1+N(w))\big(a\rho a^\dag-\frac{1}{2}\{a^\dag a,\rho\}\big)+\Gamma N(w)\big(a^\dag\rho a-\frac{1}{2}\{aa^\dag,\rho\}\big)\ .
\end{equation}
Considering a modulation of the level splitting $w$, we have for the average occupation number
\begin{equation}
\frac{d}{dt}\langle a^{\dag}a\rangle=\Gamma(N(w)-\langle a^\dag a\rangle)
\end{equation}
Thus the metric for the single control parameter $w(t)$ is in this case simply
\begin{equation}
m_{ww}=\frac{1}{\Gamma}\ .
\end{equation}
Note that a frequency-controlled harmonic oscillator can be realized in a different set up, i.e.
\begin{equation}
\beta H(t)=\frac{mw(t)^2\hat{X}^2}{2}+\frac{\hat{P}^2}{2m}
\end{equation}
(to see how this control is not equivalent to \eqref{eq:optical-MME-harmonic} it is sufficient to notice $[\dot{H},H]\neq 0$ in general). The metric arising with this control is being worked out in \cite{harry-moha}.
\subsubsection{Three-level system with a bosonic bath}
\label{3-lev}
For a final example we consider here a  three-level system satisfying a detailed balance master equation. We assume  control on each of the energy levels, while keeping the basis fixed (this is motivated by the observation that creating coherence does not increase power  in the linear-response regime~\cite{Brandner2016}). Without loss of generality we can assume the ground state energy to be zero ($\beta E_0\equiv 0$) and the two excited states with energies $\beta^{-1}{E_1(t)}\leq \beta^{-1}E_2(t)$, hence the thermal state being characterized by a probability vector
\begin{equation}
\vec{\omega}(t)=\frac{1}{1+e^{-E_1}+e^{-E_2}}
\begin{pmatrix}
1 \\ e^{-E_1} \\ e^{-E_2}
\end{pmatrix}
\end{equation} 
The evolution of the probability vector is described by the Markovian master equation~\cite{breuer-petruccione}
\begin{equation}
\label{det_bal1}
\dot{p}_i=\sum_j\Gamma_{ij}p_j\ .
\end{equation}
The rate matrix $\Gamma_{ij}$ can be obtained from \eqref{Lindmic}, yielding
\begin{align}
\label{det_bal2}
&\Gamma_{i<j}=\Gamma (N(E_j-E_i)+1)(E_j-E_i)^\alpha \nonumber\\ &\Gamma_{i>j}=\Gamma N(E_j-E_i)(E_j-E_i)^\alpha
\nonumber\\
&\Gamma_{ii}=-\sum_{j\neq i} \Gamma_{ji}
\end{align}
where $\Gamma$ is a rate, $N(w)=1/(e^w-1)$ for bosonic baths, and $\alpha$ defines the spectral density of the bath. Note that these dynamics cannot be described by the simple exponential decay \eqref{eq:multiple_relax}, so that we need to use the more general approach \eqref{mgeneral}  to find the metric. Indeed,
we use Eq.\eqref{mgeneral}, with a simplified version of $\mathfrak{J}_\omega[\dot{G}]$  due to commuting operators inside it (i.e. in this case Eq.\eqref{def:J} corresponds to $-\int_0^1 dx \hspace{1mm} \omega^{1-x}(\dot{G}-\tr(\omega \dot{G}))\omega^x=-\dot{G}\omega+\omega\Tr[\dot{G}\omega]$), obtaining
\begin{equation}
\label{3-lev_mij}
\Sigma=\sum_{ij}\dot{E_i} m_{ij} \dot{E_j} \quad \text{with} \quad m_{ij}=-\sum_k\tilde{\Gamma}^{-1}_{ik}\partial_j\omega_k
\end{equation}
where $\tilde{\Gamma}^{-1}_{ij}$ is the Drazin inverse of ${\Gamma}_{ij}$ (see e.g. \cite{crooks2018drazin,marti2018thermodynamiclength,abiuso_n-M,slowdriving} for details on the Drazin inverse) while $\partial_j\omega_k$ is the variation of $\omega_k$ due to $\dot{E_j}$, i.e. $\partial_j\omega_k=-\omega_k\delta_{jk}+\omega_j \omega_k$.
The analytic form of $m_{ij}$ can be computed from this expression even for larger systems with more than 3 levels using symbolic computation software.

\section{Generalisation of the main result: Optimal cycles for general metrics and asymmetric dissipations}
\label{SecApp:GenMainResult}

We now have the necessary tools to generalise the optimisation to more generic metrics and thermodynamic processes. In the most general case, we can consider the Carnot cycle where the cold isotherm is characterised by the control parameters $\{\lambda_j^{(c)}(s) \}$ and the hot one by $\{\lambda_j^{(h)}(s) \}$ in adimensional time units (recall $s\in [0,1]$), as well as the durations $\tau_c,\tau_h$. For time-reversal cycles considered in the main text,  $\{\lambda_j^{(c)}(s) \}=\{\lambda_j^{(h)}(1-s) \}$  (but remind that in general $\tau_c\neq\tau_h$).

We first note that under optimization of a very general class of processes where  $m_{ij}^{(h)}$ and $m_{ij}^{(c)}$ are simply proportional to each other (this class includes e.g. all the standard microscopic scenarios with same spectral density of the baths, as showed in \ref{sec:spectral_dens}), the time-reversal property is not an assumption but simply follows from the optimization process \cite{slowdriving}. This implies, as we show in the following \ref{sec:proportional_diss}, that the relevant figure of merit to be maximized is still $\Delta S^2/\Sigma$.
Therefore, we show how to accomplish this task for different structures of the metric $m_{ij}^{h,c}$, up to the most general case where the dependence on the temperature, spectral density, etc. is encoded in $\mathcal{L}^{-1}$ in the expression \eqref{mgeneral}.

Finally in Section \ref{app:asym} we will consider generalizations to cases where $m_{ij}^{(h)}$ and $m_{ij}^{(c)}$ differ significantly, so that $\{\lambda_j^{(c)}(s) \}\neq\{\lambda_j^{(h)}(1-s) \}$.

\subsection{Simple asymmetric dissipations}
\label{sec:proportional_diss}
We consider here cases in which the dissipations along the hot isotherm  are proportional to the dissipations along the cold isotherm via a constant factor that does not depend on the specific control protocol:
\begin{equation}
\Sigma_c=\sigma \Sigma_h\equiv\sigma\Sigma\ ,
\label{sigmaprophc}
\end{equation}
where $\sigma$ is a number.
 Importantly this class of cases includes the scenario where the two baths have the same (non-flat) spectral density, as explained in Sec. \ref{sec:spectral_dens} and in \cite{slowdriving}, where 
%in fact, from \eqref{relationLind}, we have that $\mathcal{L}^{(c)} = \lambda^{-\alpha} \mathcal{L}^{(h)}$ where $\mathcal{L}^{(c)} $  ($\mathcal{L}^{(h)} $) is the Lidnbladian associated to the cold (hot) bath, and $\alpha$ measures the degree of Ohmicity.  Similarly, we obtain from \eqref{mgeneral} that $m^{(c)}_{ij} =  \lambda^{\alpha} m^{(h)}_{ij} $, where $m^{(c)}_{ij}$ ($m^{(h)}_{ij}$) is the metric associated to the dissipation of the chold (hot)  bath.  Then, from \eqref{relationLind}  we have that $\mathcal{L}^{(c)} = \lambda^{-\alpha} \mathcal{L}^{(h)} $ where $\mathcal{L}^{(c)} $  ($\mathcal{L}^{(h)} $) is the Lidnbladian associated to the cold (hot) bath, and $\alpha$ measures the degree of Ohmicity.   Similarly, 
we obtained that $m^{(c)}_{ij} =  ({T_h}/{T_c})^{\alpha} m^{(h)}_{ij} $ (here $m^{(c)}_{ij}$ ($m^{(h)}_{ij}$) is the metric associated to the dissipation of the cold (hot)  bath), which implies  $\Sigma_c =  ({T_h}/{T_c})^{\alpha} \Sigma_h $.

%In general we thus consider
%\begin{equation}
%\Sigma_c=\sigma \Sigma_h\equiv\sigma\Sigma\ ,
%\end{equation}
%where $\sigma$ is a number. 

Starting from \eqref{sigmaprophc}, the power to maximize is given by
\begin{equation}
P=\frac{(T_h-T_c)\Delta S - \Sigma(\frac{T_h}{\tau_h}+\frac{\sigma T_c}{\tau_c})}{\tau_h+\tau_c}
\end{equation}
under the efficiency constraint
\begin{equation}
1-\frac{T_c(1+\frac{\sigma \Sigma}{\Delta S \tau_c})}{T_h(1-\frac{\Sigma}{\Delta S \tau_h})}=\gamma\big(1-\frac{T_c}{T_h}\big)
\end{equation}
As the only time unit is $\Sigma/\Delta S \equiv \tau^*$ and the temperature ratio $r=\frac{T_c}{T_h}$ the problem can be rephrased as
%\begin{equation}
%\max_{\tau_c,\tau_h} \Bigg[ T_h\Delta S \frac{(1-r) - (\frac{\tau^*}{\tau_h}+\frac{r \sigma \tau^*}{\tau_c})}{\tau_h+\tau_c}\Bigg] \qquad \text{with} \qquad 1-r\frac{1+\sigma \tau^*/\tau_c}{1- \tau^*/\tau_h}=\gamma(1-r)
%\end{equation}
\begin{equation}
 \frac{T_h\Delta S}{\tau^*} \max_{x_c,x_h} \Bigg[ \frac{(1-r) - (\frac{1}{x_h}+\frac{r \sigma}{x_c})}{x_h+x_c}\Bigg] \qquad \text{with} \qquad 1-r\frac{1+\sigma/x_c}{1- 1/x_h}=\gamma(1-r)
\end{equation}
where $x_j=\tau_j/\tau^*$. Here the maximization with its constraint is expressed in full adimensional terms, meaning that after solving it the resulting power will be in the form $\frac{T_h(\Delta S)^2}{\Sigma}f(\gamma,r,\sigma)$ for some scalar function $f$. % (as a matter of a fact, the function $f$ can be analytically obtained but we do not show it here due to its length and lack of physical insights). 
% which can be then maximized as previously showed.
The general form of the function $f$ can be obtained analytically but it is in general  non-trivial  and very lengthy (so we do not show it here), but in the limit of high efficiency it simplifies to
\begin{equation}
\label{eq:pmax_simple_asym}
P^{(\rm max)}_{\gamma\approx 1}=\frac{T_h\Delta S ^2}{\Sigma}\frac{(1-r)^2}{r(1+\sqrt{\sigma})^2}(1-\gamma)+O((1-\gamma)^2) \ .
\end{equation}
At this point the optimisation of the power  boils down to the maximisation  of $(\Delta S)^2/\Sigma$, as in the main text,. 
%Given the assumption of $\sigma$ not depending on the specific protocol, after time optimization the shape dependence of the results will thus be left again only in the term $(\Delta S)^2/\Sigma$, as in the main text. 
In the following we show how to maximise $\Delta S^2/\Sigma$ in general scenarios, and prove that asymptotically infinitesimal cycles are optimal.

Before, let us note that given any reasonable figure of merit between efficiency and power, the relevant figure of merit will always be $(\Delta S)^2/\Sigma$. Given as objective any figure of merit $f(\eta ,P)$
  expressed in terms of the efficiency and the power, after optimization on
  $\tau _h,\tau _c$ the resulting value $\protect \mathaccentV
  {bar}016{f}:=f(\protect \mathaccentV {bar}016{\eta },\protect \mathaccentV
  {bar}016{P})$ (we indicate with $\protect \mathaccentV {bar}016{}$ quantities
  after time optimization) will be given, by
  $\protect \mathaccentV {bar}016{\eta }=\protect \mathaccentV {bar}016{\eta
  }(T_h,T_C,\sigma)$ and $\protect \mathaccentV {bar}016{P}=\protect \mathaccentV
  {hat}05E{P}(T_h,T_c,\sigma) (\Delta S) ^2 /\Sigma $ for some adimensional function
  $\protect \mathaccentV {bar}016{\eta }$ and homogeneous function $\protect
  \mathaccentV {hat}05E{P}$ of the temperatures, by dimensional analysis.
  Moreover, any reasonable $f$ will be monotonously increasing in both its
  parameters singularly (i.e. for a given efficiency we wish to enhance the
  power and vice-versa), therefore after speed optimization it will still be
  possible to improve the engine performance by increasing the ratio $(\Delta
  S)^2/\Sigma $. The role of the above mentioned ratio as a characteristic
  scale defining the performance of both engines and refrigerators was already
  noticed in \cite{Hernndez2015} without further analysis.

\subsubsection{Dependence of the time scale in $G$}
\label{app:changing_teq}
As a simple first extension of our results, we consider the case of the standard thermodynamic metric with a time-scale that can depend on the point of the parameter space $G=\beta H$. In other words, we consider the heuristic model of exponential thermalisation \eqref{trivialLind} with a $G$-dependent $\tau_{\rm eq}$ (see also \cite{Bonanca2014d} for a microscopic derivation). %, i.e., $\tau_{\rm eq}=\tau_{\rm eq}(G)$. %, or equivalently 
%\begin{align}
%m_{ij}= \tau_{\rm eq}(\{\lambda_j\}) \frac{\partial^2 \ln \mathcal{Z}}{\partial \lambda_i \partial \lambda_j}.
%\label{freeenergymetricII}
%\end{align}
Following the reasoning of the main text, we can define an infinite-dimensional scalar product given by
\begin{align}
\langle A, B\rangle_{\omega} \equiv \int_0^1 ds \hspace{1mm} {\rm cov}_{\omega_s}(A,B)
\end{align}
with $\omega_s = e^{-G_s}/\tr(e^{-G_s})$, and where we note that the scalar product is defined for a fixed trajectory $G_s$ with $s\in [0,1]$.  
%For symmetric protocols ($G(s\tau_c)=G(\tau_c+\tau_h(1-s))$ as in the main text), 
We then have, 
\begin{align}
\label{inequality}
\frac{(\Delta S) ^2}{\Sigma} = \frac{\left(-\int_0^1 ds \hspace{1mm} {\rm cov}_{\omega}(G,\dot{G})\right)^2}{\int_0^1 ds \hspace{1mm} {\rm cov}_{\omega}(\dot{G},\dot{G}) \tau_{\rm eq}}= \frac{\left(\langle \tau^{-1/2}G,\tau^{1/2}\dot{G}\rangle_\omega\right)^2}{\langle \tau^{1/2}\dot{G},\tau^{1/2}\dot{G}\rangle_{\omega}}\leq \langle \tau^{-1/2}G,\tau^{-1/2}G\rangle_\omega = \int_0^1 ds \hspace{1mm} \frac{\mathcal{C}}{\tau_{\rm eq}},
\end{align}
so we conclude that the results of the main text can be naturally extended to time-scales that depend on $\beta H$, 
 meaning that the performance is upper-bounded by small cycles with proportional modulation of the Hamiltonian $\dot{H}\propto H$ performed on the point where the ratio between heat capacity and relaxation time of the system is maximum.

%The results here presented are actually a special case of the following.

\subsubsection{General metrics}
\label{app:C-S}
Let us consider the maximisation of $\Delta S^2/\Sigma$ for more general protocols where the thermodynamic metric is given by the general expression \eqref{mgeneral} which depends on the particular Lindbladian describing the thermalisation dynamics. 
%We assume that:
%\begin{itemize}
%\item The protocol is time-reversal symmetric, i.e. it satisfies $G(s\tau_c)=G(\tau_c+\tau_h(1-s))$ as in the main text. This property, under optimization, is actually a consequence of the following assumption:
%\item The dissipation metrics $m_{ij}$ \eqref{mgeneral} is the same along the cold and hot isotherm.
%\end{itemize}   
%In this case, note that the Lindbladian enters into the metric through the Drazin inverse $\mathcal{L}^{-1}$. We further assume that both baths have the same spectral density.
Following the  considerations of the main text, to maximise
% In this case the dissipation $\Sigma$ in \eqref{SigmaslowdrivingIII} will depend on the particular bath, i.e., 
%\begin{align}
%$\Sigma_{x}= \sum_{ij} \int_0^1 {\rm d}t' \hspace{1mm}\dot{\lambda}_i m^{(x)}_{ij} \dot{\lambda}_j$, $x=h,c$. 
%\end{align}
%In this subsection we make the assumption $m^{(c)}_{ij} = f(\beta,J(w)) m^{(h)}_{ij} $ where $f(\beta,J(w))$ is a function that depends on the temperature and the spectral function of the bath. This assumption holds true when: both baths have the same spectral density, and each bath is in the low/high temperature regime as described in Sec. \cite{Sec:temperature}.  Indeed, in this case the metric becomes proportional to $\beta$ (in the high temperature regime) and is independent of $\beta$ (in the low temperature regime). 
  $(\Delta S)^2/\Sigma$, we can expand them as
 \begin{equation}
 \label{smexpg}
 G_{s} = \sum_j \lambda_j(s) X_j \quad \Rightarrow\quad  \Delta S =\sum_i \int_0^1 s_i\dot{\lambda}_i
 \end{equation}
 with $s_i=\Tr[(\Tr[G\omega]\omega-G\omega)X_i]$, and
\begin{align}
\Sigma = \sum_{ij} \int_0^1 {\rm d}t \hspace{1mm}\dot{\lambda}_i m_{ij} \dot{\lambda}_j.
\label{SigmaslowdrivingII}
\end{align}
In contrast to the main text, where $m_{ij}$ is given by Eq.~(4) in the main text, here we a general metric of the form \eqref{mgeneral}. The Cauchy-Schwarz inequality can be applied by considering the following:
   %, or a scalar product, which in turn gives the notion of a thermodynamic length.  
%In fact, the concept of thermodynamic length gives a procedure for minimisng $\Sigma$ between two fixed points  $G_0$ and $G_1$, by finding the geodesics associated to the metric $m$ in the thermodynamic control space $\lambda(t')$ \cite{marti2018thermodynamiclength}. In this case, since $\Delta S$ is fixed by the two endpoints  $G_0$ and $G_1$, maximizing ${(\Delta S)^2}/{\Sigma}$ is equivalent to minimizing $\Sigma$. Here, we instead assume that the  $\lambda$ can be chosen arbitrarily within the space of allowed transformations, and find the optimal cycle that minimises  ${(\Delta S)^2}/{\Sigma}$. 
\paragraph*{Inequality.} Consider two vectors $\vec{a},\vec{b}$ and a quadratic invertible form $g>0$ defined on their vector space.
Then the standard C-S inequality applied to $g^{1/2}\vec{a}$ and $g^{-1/2}\vec{b}$ tells
\begin{equation}
(\vec{a}\cdot\vec{b})^2=\big((g^{1/2}\vec{a})\cdot(g^{-1/2}\vec{b})\big)^2\leq |g^{1/2}\vec{a}|^2|g^{-1/2}\vec{b}|^2=(\vec{a}^Tg\vec{a})(\vec{b}^Tg^{-1}\vec{b})\ .
\end{equation}
If we now consider this inequality applied to two vectors
\begin{equation}
s_j(t),\ \lambda_j(t)\qquad j=1,\dots,k\quad 0\leq t\leq 1\ ,
\end{equation}
and the metric
\begin{equation}
g_{it,jt'}=m_{ij}(t)\delta(t-t') \ ,
\end{equation}
where $m(t)$ is a positive time-dependent quadratic form in $\mathbb{R}^k$ , we have 
\begin{equation}
g^{-1}=m^{-1}_{ij}(t)\delta(t-t')\ ,
\end{equation}
and thus it is possible to write the C-S inequality as
\begin{equation}
\bigg(\int_0^1 s_j(t)\lambda_j(t)\bigg)^2\leq \bigg(\int_0^1 s_i(t)m_{ij}^{-1}(t)s_j(t)\bigg)\bigg(\int_0^1 \lambda_i(t)m_{ij}(t)\lambda_j(t)\bigg)\ .
\end{equation}

This inequality allows us to  bound the figure of merit for the maximum power as
\begin{equation}
\label{eq:maxpow_generic}
\frac{(\Delta S) ^2}{\Sigma} \leq \max_{\lambda} \sum_{ij} s_i m_{ij}^{-1}s_j\ .
\end{equation}
where we recall that $s_i=\Tr[(\Tr[G\omega]\omega-G\omega)X_i]$, and the $X_i$ are given by the expansion \eqref{smexpg}. Hence, the initial optimisation over all possible trajectories $\{ \lambda_j(s) \}$  boils down to maximising a single real function given the control parameters. In other words,  our result shows that in order to design optimal finite-time Carnot cycles one does not need to optimise over all trajectories $\{ \lambda_j(s)\}$ with $s \in [0,1]$, but it is enough to search a suitable point in the thermodynamic space $\{ \lambda_j\}$. 
To saturate it in practice, one needs to maximise  $\vec{s} m^{-1} \vec{s}$ over the control parameters~$\vec{\lambda}$, 
\begin{align}
\lambda^*={\rm argmax}\left(\sum_{ij} s_i m_{ij}^{-1}s_j\right),
\end{align}
and consider infinitesimal variations around this optimal point: 
\begin{equation}
\label{eq:optimal_traj}
\vec{\lambda}=\vec{\lambda}^*+\epsilon t \vec{\mu}, \quad\epsilon \ll 1, \quad \vec{\mu}= m^{-1} \vec{s}\ ,
\end{equation} 
where we have used the more compact vector notation: $\vec{s}=\{s_1, s_2,... \}$, etc.
The direction of the modulations is defined, from the Cauchy-Schwarz saturation condition, by the vector $ m^{-1} \vec{s}$, while the normalization of the modulation is in principle to be taken infinitesimal. The abstract expression for $\vec{\mu}$ corresponds, in the case of the analysed in the main text, to $\vec{\lambda}^*$ itself, i.e. the modulations of the control parameters are proportional to the control themselves in that case.
%It is in this sense that our approach significantly simplifies the design of optimal finite-time (Carnot) cycles. 
\subsubsection{Examples}

\begin{figure}
\includegraphics[width=0.48\textwidth]{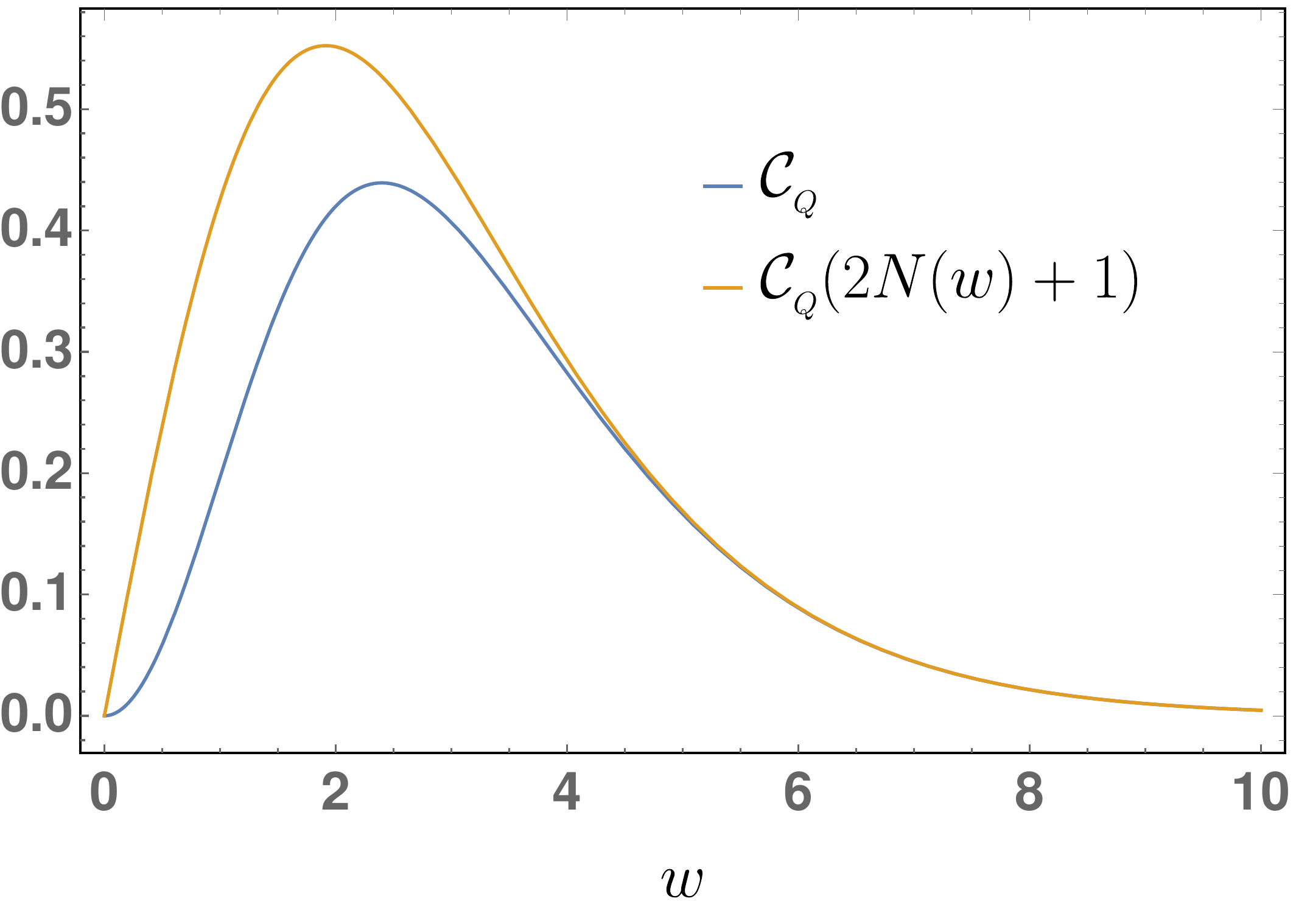} 
\caption{Maximization, as  function of the gap $w$ of the (adimensional) heat capacity of a Qubit divided by the thermalization scale $\tau_{\rm eq}\equiv \Gamma^{-1}(2N(w)+1)^{-1} $, in units of $\Gamma$. A comparison can be made with the maximization of the heat capacity alone.
}
\label{fig:qbit}
\end{figure}

\paragraph{\textbf{Qubit with optical master equation.}}
We first consider here a slowly driven qubit with full hamiltonian control (here $G$ is the adimensional Hamiltonian in temperature energy units)
\begin{equation}
G(t)=\frac{w(t)}{2}\hat{n}(t)\cdot\vec{\sigma}
\end{equation}
in contact with a bosonic bath, so that the evolution is described by the adiabatic master equation (in the interaction picture) %and dynamical equation
\begin{equation}
\dot{\rho}=\Gamma(1+N(w))\big(\sigma^-\rho\sigma^+-\frac{1}{2}\{\sigma^+\sigma^-,\rho\}\big)+\Gamma N(w)\big(\sigma^+\rho\sigma^--\frac{1}{2}\{\sigma^-\sigma^+,\rho\}\big)\
\end{equation}
Here, $\sigma^\pm\equiv \sigma^\pm(t)=\sigma^x(t)\pm\sigma^y(t)$ in the basis where $\sigma^z(t)=\hat{n}(t)\cdot\vec{\sigma}$, and $\omega\equiv \omega(t)$, as in Section~\ref{subsec:qbit_mij}. 
Given the metric~\eqref{eq:optical_qubit_mij} and considering that the variation of entropy is non-zero only along the instantaneous $z$ direction $\hat{n}(t)$ it is evident how, to maximize $\Delta S/\Sigma^2$, changing the direction of $\hat{n}$ is useless (as noted in \cite{marti2018thermodynamiclength}, generation of coherences is detrimental, and the eigenvectors of $m_{ij}$ \eqref{eq:optical_qubit_mij} are larger along the tangential directions $x$ and $y$), hence the optimal trajectories can be recognized to be in the form of simple gap modulation $w(t)$, reducing the metric to
\begin{equation}
m_{ww}=\frac{1}{\Gamma (2N(w)+1)}{\rm cov}(\sigma^z,\sigma^z)\ , \quad \text{implying} \quad \Sigma=\int \frac{{\rm cov}(\dot{G},\dot{G})}{\Gamma(2N(w)+1)}\ .
\end{equation}
Therefore we fall into the category of systems analysed in Section~\ref{app:changing_teq}, for which the optimal control consists in modulations on the point where the ratio between the heat capacity ${\rm Var}_\omega(G)$ and the relaxation timescale is maximum. In this case $\tau_{\rm eq}\equiv \Gamma^{-1}(2N(w)+1)^{-1} $ and $\mathcal{C}_Q=w^2e^{-w}/(1+e^{-w})-(we^{-w}/(1+e^{-w}))^2 $. In Fig.~\ref{fig:qbit} we show the maximization of $\mathcal{C}_Q\Gamma(2N(w)+1)$, corresponding to the maximum value of $(\Delta S)^2/\Sigma$ in this case, choosing $\Gamma$ constant (i.e. flat spectral density, $\alpha=0$). Allowing $\alpha\neq 0$ would not change significantly the difficulty of the maximization problem, here $\alpha=0$ has been chosen to allow a comparison with the case studied in the main text, where the thermalization timescale does not depend on the value of the Hamiltonian.
To sum up, by making use of equation \eqref{eq:pmax_simple_asym}, the full maximization of power for a qubit in contact with bosonic thermal baths, leads to
\begin{equation}
\label{res:optical_qubit_maxpow}
P^{(\rm max)}_{\gamma\approx 1}=\max_w \{C_Qw^\alpha(2N(w)+1)\}\frac{\Gamma(T_h-T_c)^2}{T_c(1+\sqrt{(T_h/T_c)^\alpha})^2}(1-\gamma)+O((1-\gamma)^2) 
\end{equation}
where the maximization result depends on $\alpha$.

\paragraph{\textbf{Harmonic oscillator}}
\begin{figure}
\includegraphics[width=0.46\textwidth]{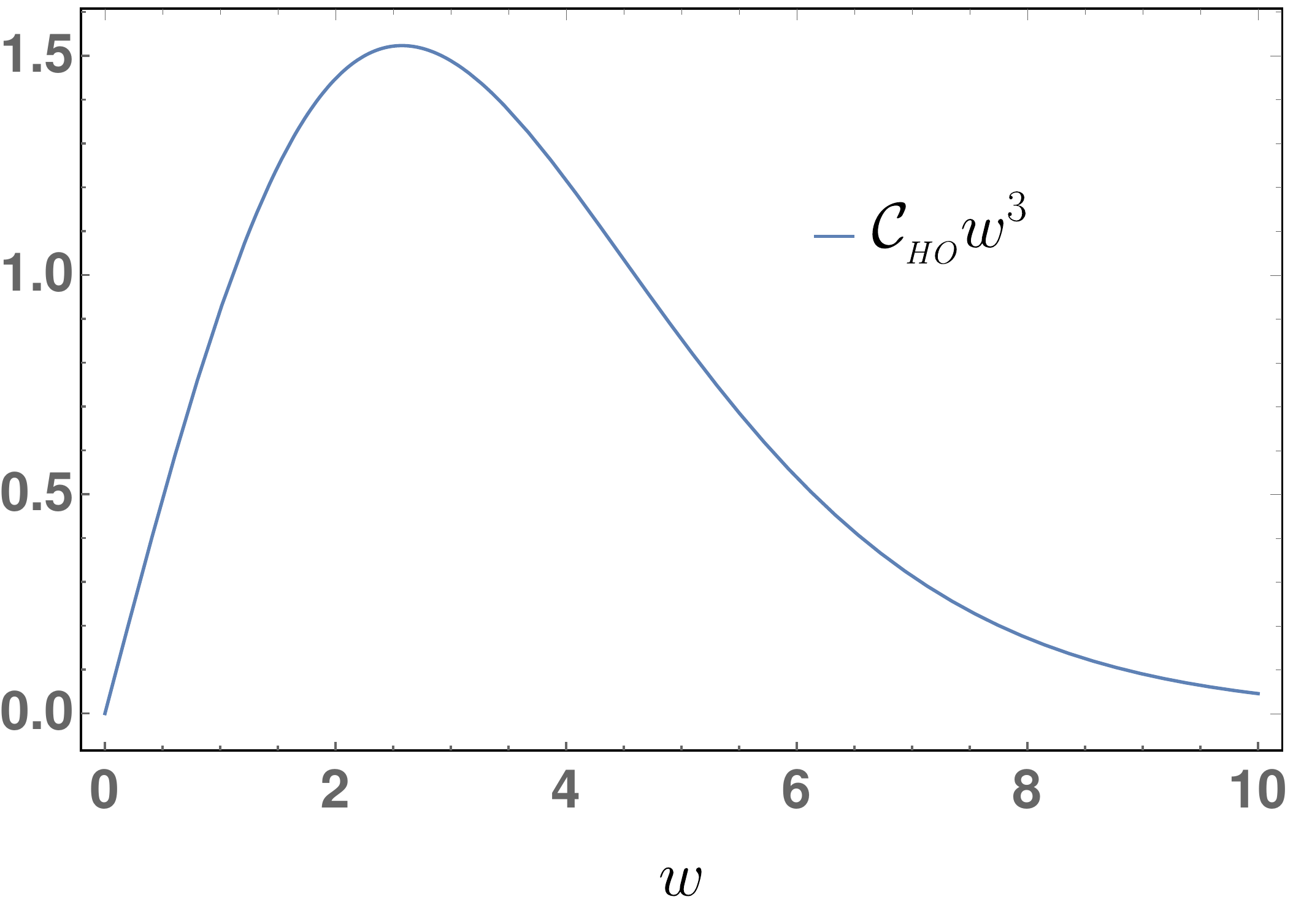}
\caption{Similar maximization for an harmonic oscillator attached to a bath of spectral density $\alpha=3$.}
\label{fig:harmosc}
\end{figure}
The above analysis for the qubit can be replicated thoroughly for the controlled harmonic oscillator described in \ref{simple_harmosc_metric}, and the final result 
\begin{equation}
P^{(\rm max)}_{\gamma\approx 1}=\max_w \{C_{HO}w^\alpha\}\frac{\Gamma(T_h-T_c)^2}{T_c(1+\sqrt{(T_h/T_c)^\alpha})^2}(1-\gamma)+O((1-\gamma)^2) \ .
\end{equation} 
where heat capacity of the harmonic oscillator is equal to $\mathcal{C}_{HO}=(2\sinh(w/2))^{-2}$. In Fig.~\ref{fig:harmosc} an example is shown for $\alpha=3$.

\begin{figure}
\includegraphics[width=0.4\textwidth]{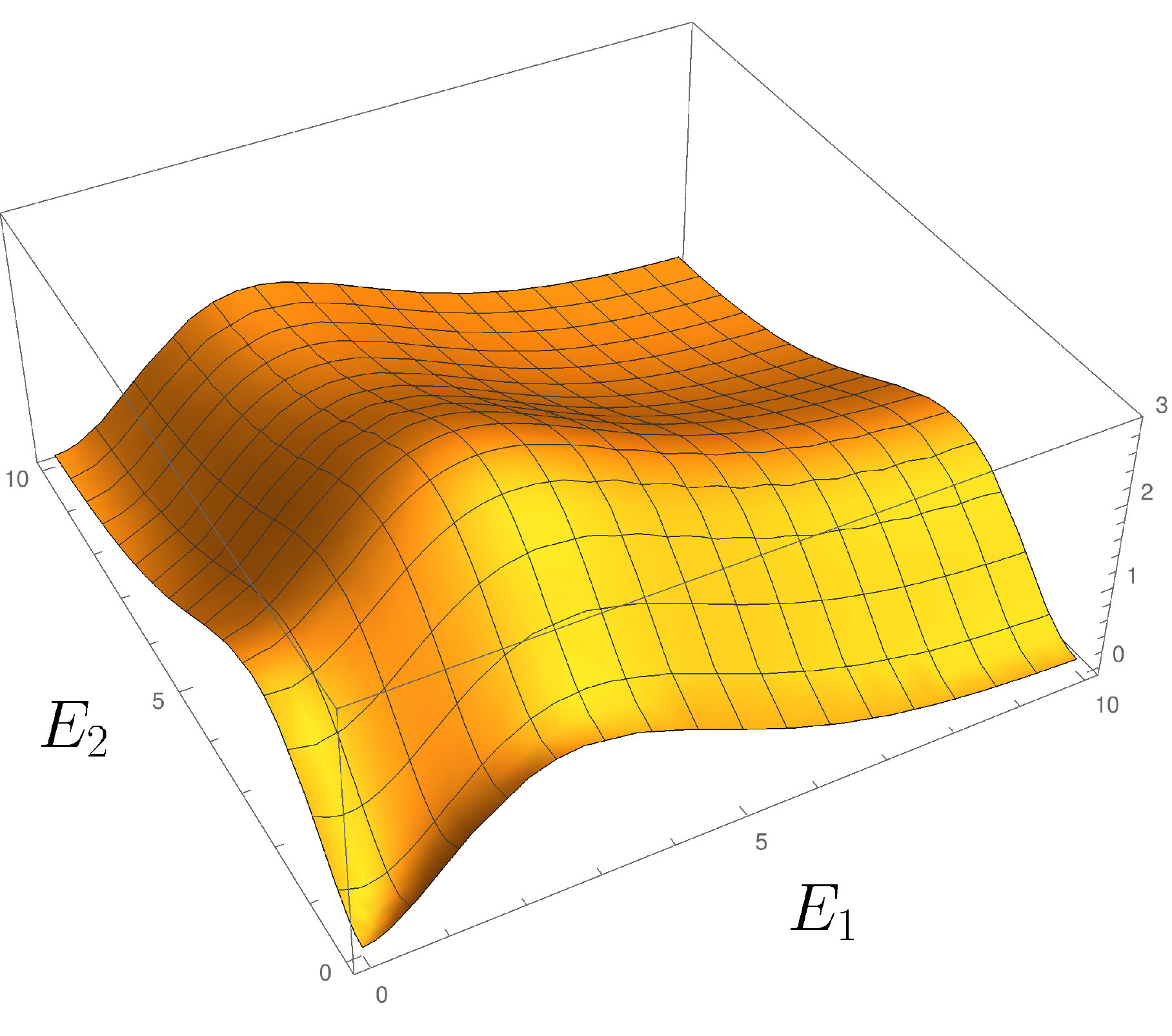} \ \
\includegraphics[width=0.55\textwidth]{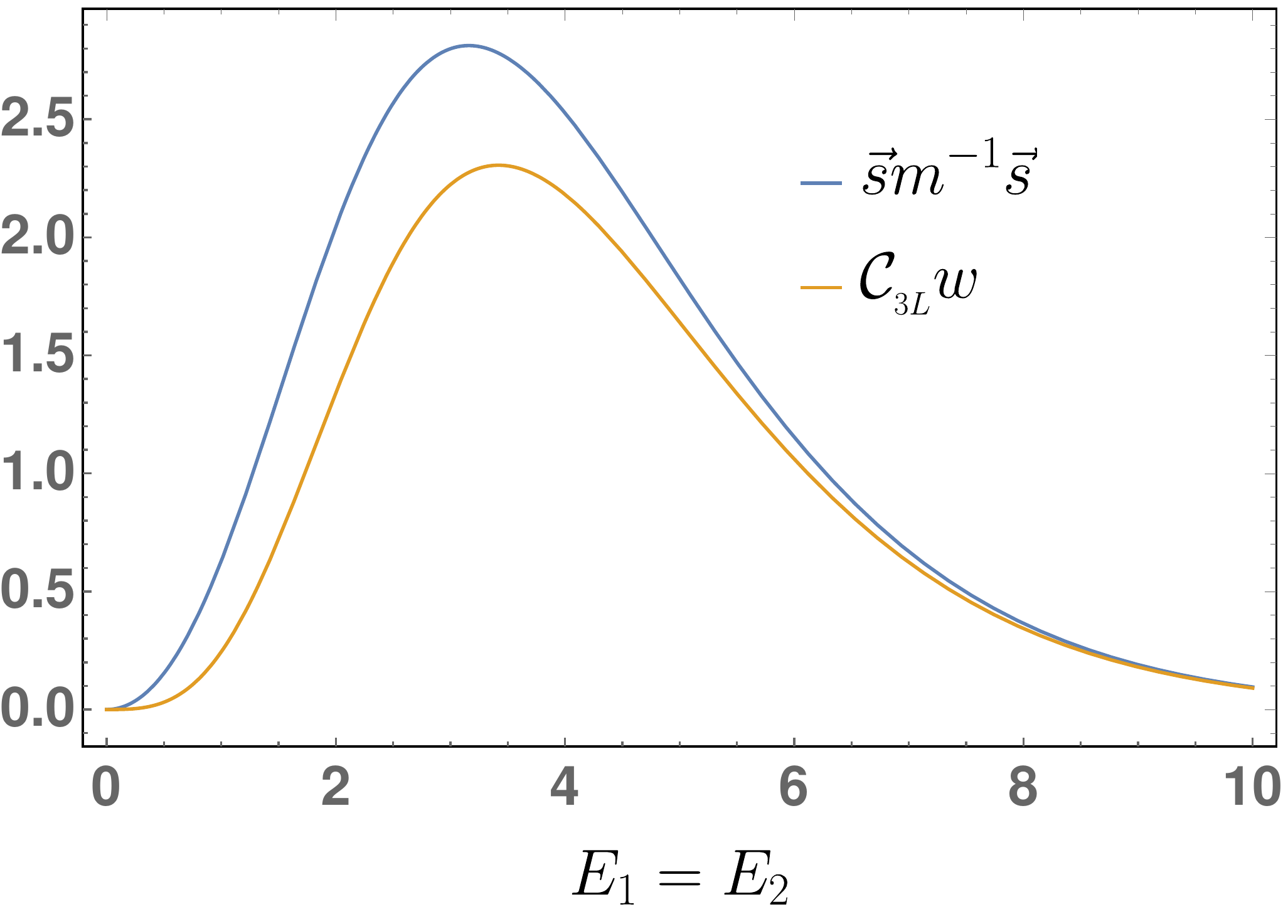}
\caption{(Left) The value of  $\sum_{ij}s_i m_{ij}^{-1} s_j$ (which bounds $(\Delta S)^2/\Sigma$, as shown in \eqref{eq:maxpow_generic}), for the 3-level system described in \ref{3-lev}, with $\alpha=1$.
(Right) Plot of the bound compared to the maximization of the heat capacity multiplied by the rate $\Gamma w$ (in units of $\Gamma$), on the bisector $E_1=E_2$.}
\label{fig:3lev}
\end{figure}

\paragraph{\textbf{3-level system with master equation}}
Finally we consider a 3-level system as described in \ref{3-lev}, controlled by the energy levels of the two excited states $E_1$ and $E_2$, and whose evolution is given by a standard detailed balanced master equation \eqref{det_bal1}-\eqref{det_bal2}. By following the general approach described in \ref{app:C-S}, we  bound $\Delta S^2/\Sigma$ by  $\sum_{ij} s_i m_{ij}^{-1} s_j$, see \eqref{eq:maxpow_generic}. We  symbolically compute the metric \eqref{3-lev_mij} and the vector $\{ s_i \}$ (see details in Sec. \ref{3-lev})  in order to express  $\sum_{ij} s_i m_{ij}^{-1} s_j$ as a function of $\{E_{1},E_{2}\}$, and then maximise it over $\{E_{1},E_{2}\}$ numerically. An example is shown in Fig.~\ref{fig:3lev}, %As an example we chose the spectral density of the baths to be linear $\alpha=1$, and in  Fig.~\ref{fig:3lev}
where we plot $\sum_{ij} s_i m_{ij}^{-1} s_j$ for a linear spectral density $\alpha=1$. Note that the maximum is obtained for $E_1=E_2$. We also find that  in this case the vector $\vec{\mu}$ \eqref{eq:optimal_traj} is proportional to $\begin{pmatrix} 1 \\ 1 \end{pmatrix}$ on the bisector, meaning the modulation on the optimal working point are symmetric on $E_1$ and $E_2$.
The maximum power results in such a case
\begin{equation}
P^{(\rm max)}_{\gamma\approx 1}\approx 2.8\frac{\Gamma(T_h-T_c)^2}{T_c(1+\sqrt{(T_h/T_c)})^2}(1-\gamma)+O((1-\gamma)^2) \ .
\end{equation} 

Furthermore, in Fig. 4 (right), we also compare
 $\sum_{ij} s_i m_{ij}^{-1} s_j$  with the heat capacity $\mathcal{C}$ divided by the characteristic timescale $\tau_{\rm eq}=\Gamma^{-1} w^{-1}$ (but remember the thermalization process is not simply exponential in this case), and it can be seen how this ratio  provides a rather good approximation of the
maximal power. That is, while $\mathcal{C}/\tau_{\rm eq}$  is not exactly the correct figure of merit in this case, it can be used
as a relevant figure of merit to determine where the optimal power is. This example also illustrates that our approach can be used to deal with more complex dissipative dynamics than the standard exponential decays discussed in the main text.

\subsection{General asymmetric protocols}
\label{app:asym}
Finally, we consider the most general setting (time-reversal asymmetric protocol, different baths, generic dependence on the temperature of $\mathcal{L}$):  $\Sigma_c\neq\Sigma_h$. Then, the maximum power expression after isotherms-duration tuning (with no efficiency constraints) is given by \cite{ma2018universal}
\begin{align}
\label{maxPowasym}
P^{(max)}=\frac{(\Delta S)^2}{\big(\sqrt{T_c \Sigma_c}+\sqrt{T_h \Sigma_h}\big)^2}\frac{(T_h-T_c)^2}{4} 
\end{align} which can be thus written as from Eq.s~\eqref{sigmageneral}-\eqref{mgeneral}
\begin{equation}
\label{eq:asymm}
\frac{(\sum_j \int_0^1 {\rm d} t \hspace{1mm}  s_j \dot{\lambda}_j )^2}{\Big(\sqrt{T_c\sum_{ij} \int_0^1 {\rm d}t \hspace{1mm}\dot{\lambda}_i m^{(c)}_{ij} \dot{\lambda}_j}+\sqrt{T_h\int_0^1 {\rm d}t \hspace{1mm}\dot{\lambda}_i m^{(h)}_{ij} \dot{\lambda}_j}\Big)^2}\frac{(T_h-T_c)^2}{4}\ ,
\end{equation}
where now two different metrics $m^{(j)}$ appear in the denominator. In this case we lose the simple structure (scalar product$)^2/($quadratic form), hence it is not possible to apply directly the C-S inequality; nevertheless it is possible to upper-bound Eq.~\eqref{eq:asymm} using simple inequalities e.g.
\begin{align}
\label{ineq:asymm1}
\eqref{eq:asymm}& \leq \max_{x=c;h} \Bigg[ \frac{(\sum_j \int_0^1 {\rm d} t \hspace{1mm}  s_j \dot{\lambda}_j )^2}{4T_x\sum_{ij} \int_0^1 {\rm d}t \hspace{1mm}\dot{\lambda}_i m^{(x)}_{ij} \dot{\lambda}_j}\Bigg]\frac{(T_h-T_c)^2}{4} \qquad &\text{using }& \sqrt{a}+\sqrt{b}\geq 2\min (\sqrt{a},\sqrt{b})\ ;&\\
\label{ineq:asymm2}
\eqref{eq:asymm}& \leq\frac{(\sum_j \int_0^1 {\rm d} t \hspace{1mm}  s_j \dot{\lambda}_j )^2}{\sum_{ij} \int_0^1 {\rm d}t \hspace{1mm}\dot{\lambda}_i [T_c m^{(c)}+T_h m^{(h)}]_{ij} \dot{\lambda}_j}\frac{(T_h-T_c)^2}{4} \qquad &\text{using }& \sqrt{a}+\sqrt{b}\geq \sqrt{a+b}\ .
\end{align}
These can be now optimized using C-S as we did in the previous section, and depending on the relative size (in the interval $\frac{1}{3}\leq \frac{a}{b}\leq 3$ the former is tighter, otherwise the latter) they will give a bound (not tight in general).
Note that while inequalities (\ref{ineq:asymm1}-\ref{ineq:asymm2}) are useful to give upperbounds to the maximum power theoretically obtainable, in practical terms it is also possible to give lower bounds, which are useful to certify that it is possible to reach \emph{at least} a given value of the power, and maximizing it by the same methods we used in the main work. Namely the bounds can be written 
\begin{align}
\label{ineq:asymm3}
\eqref{eq:asymm}& \geq \min_{x=c;h} \Bigg[ \frac{(\sum_j \int_0^1 {\rm d} t \hspace{1mm}  s_j \dot{\lambda}_j )^2}{4T_x\sum_{ij} \int_0^1 {\rm d}t \hspace{1mm}\dot{\lambda}_i m^{(x)}_{ij} \dot{\lambda}_j}\Bigg]\frac{(T_h-T_c)^2}{4} \qquad &\text{using } & \sqrt{a}+\sqrt{b}\leq 2\max (\sqrt{a},\sqrt{b})\ ,\\
\label{ineq:asymm4}
\eqref{eq:asymm}& \geq\frac{(\sum_j \int_0^1 {\rm d} t \hspace{1mm}  s_j \dot{\lambda}_j )^2}{\sqrt{2}\sum_{ij} \int_0^1 {\rm d}t \hspace{1mm}\dot{\lambda}_i [T_c m^{(c)}+T_h m^{(h)}]_{ij} \dot{\lambda}_j}\frac{(T_h-T_c)^2}{4} &\qquad \text{using } &\sqrt{a}+\sqrt{b}\leq \sqrt{2(a+b)}\ .
\end{align}
As a final observation, we note how all of the above can be applied also to the power in the high efficiency regime, which is (cf. Eq.~\eqref{eq:pmax_simple_asym})
\begin{equation}
P^{(\rm max)}_{\gamma\approx 1}=\frac{{\Delta S ^2}({T_h}-{T_c})^2}{T_c(\sqrt{\Sigma_h}+\sqrt{\Sigma_c})^2}(1-\gamma)+O((1-\gamma)^2) \ .
\end{equation}

\section{Maximizing the heat capacity vs. degree of control}
\label{sec:heatcap_models}
The optimal cycles described in the main text (induced by the standard thermodynamic metric (4)) make the maximal power proportional to the heat capacity  \eqref{maxpower} of the chosen working point of the infinitesimal Carnot cycle. The task of maximizing the power is then translated in finding the point in the control parameter space that maximizes the heat capacity, i.e. the variance of the adimensional Hamiltonian $G$ for a thermal state $\omega=e^{-G}/\Tr[e^-G]$. We show in the following 3 paradigmatic cases that are summed up in Fig.\ref{fig:supra-extensive} showing how different degrees of control offer different possible performance on the same system, which is made of $N$ qubits. 

{\bf (a) Full control over the spectrum}. We first assume total control on the Hamiltonian of the $N$ qubits, which is made up of  $D=2^N$ energy levels. That is, any desired (possibly long-range) interaction can be engineered, so that the $D$-level spectrum  can be controlled at will. While this might be extremely challenging to realise in practice, it is useful to consider this situation to obtain a fundamental upper bound on the maximal power.   Indeed, the maximization of heat capacity $\mathcal{C}$ of a general $D$-dimensional system at thermal equilibrium has been carried out in~\citep{Reeb2014,Correa-Mehboudi_optimalthermalprobes}. The optimal Hamiltonian consists of a ground level and a $D-1$ degenerate level, with an optimal gap $x$ in adimensional units (i.e. rescaled by the temperature) defined by $e^x=(D-1)(x+2)/(x-2)$ and the corresponding $\mathcal{C}$ is $\mathcal{C}_{\rm max}=x^2e^x (D-1)/(D-1+e^x)^2$.
%\begin{align}
%\mathcal{C}_{\rm max}=x^2e^x \frac{D-1}{(D-1+e^x)^2}.
%\end{align} 
This expression gives in the asymptotic regime ($D\rightarrow\infty$) $x \simeq \ln D$, hence $\mathcal{C}_{\rm max} \simeq (\ln  D)^2/4$; which in terms of the particle number  $N\propto\ln D$ means $\mathcal{C}_{\rm max} \simeq N^2/4$ for $N\gg 1$, i.e., a quadratic scaling. 

\emph{On the robustness of result {(a)}.} Suppose the control of the optimal energy gap $x$ has some imprecision (or simply needs to be modulated to perform the cycle). We show here that as far as the modulation/imprecision doesn't scale with the dimension, the heat capacity behaves smoothly. Suppose indeed that $x=(1+\epsilon)\ln(D-1)$. Then the adimensional variance of the flat Hamiltonian with $d\equiv D-1$ degenerate excited states and gap $x$ is
\begin{equation}
\label{eq:robustness_degmod}
\frac{de^{-x}x^2}{1+de^{-x}}-\frac{d^2e^{-2x}x^2}{(1+de^{-x})^2}=(\ln d)^2(1+\epsilon)^2\Big(\frac{d^{-\epsilon}}{1+d^{-\epsilon}}-\frac{d^{-2\epsilon}}{(1+d^{-\epsilon})^2}\Big)=(\ln d)^2\frac{(1+\epsilon)^2}{4\cosh^2(\frac{\epsilon}{2}\ln d)}
\end{equation}
which, to keep the scaling as $(\ln d)^2\sim N^2$ needs the rest to stay finite, i.e. $\epsilon$ to scale as $1/\ln d=1/N$.

{\bf  (b) N independent qubits}. As another extreme case, corresponding to almost no control on the spectrum, we can consider $N$ independent qubits, that is an Hamiltonian $H^{(N)}=\sum_{i=1}^N \lambda_i(t)\sigma^z_i$. Note that given the qubits do not interact the thermal heat capacity will result additive, and the optimal gap will be the same for all the qubits $\lambda_i=\lambda_j$. This means it is enough to solve  the previous case for $D=2$ (i.e. $N=1$), to find $C_{\rm max}\approx 0.439 N$ (with an optimal gap $\lambda^*_i\approx 2.40$), in agreement with \cite{abiuso_n-M}. %This result should be compared with the scaling $\mathcal{C}_{\rm max} \propto N^2$ obtained for full control on the spectrum. 
%In case the working fluid consists of a two-level system with adimensional gap $\beta E=g$, the thermal state is characterized by the ground state Gibbs probability $p=\frac{e^g}{1+e^g}$ (while the excited state has population $1-p=\frac{1}{1+e^g}$) so that $g=\ln(p/(1-p))$. Hence the variance of the adimensional Hamiltonian can easily be written as
%\begin{equation}
%\langle G^2\rangle -\langle G\rangle^2=(p-p^2)\big[\ln\big(p/(1-p)\big)\big]^2
%\end{equation}
%which is the result found in \cite{abiuso_n-M}, and has a maximum $\sim 0.11$ for $p\sim 0.92$. 

{\bf  (c) Ising chain}. Finally, we consider an Ising chain ($H^{(N)}=-\lambda_1(t)\sum_{i=1}^N \sigma^z_i\sigma^z_{i+1}-\lambda_2(t)\sum_{i=1}^N\sigma^z_i$), and assume control over $\lambda_{1,2}(t)$. Numerical results  in Fig.   \ref{fig:supra-extensive} show how the interactions  allow for substantially improving on (b), although asymptotically we obtain a linear scaling in~$N$, $C_{\rm max} \simeq 0.59 N$ for $N\gg 1$. This linear scaling is in fact expected for systems away of a phase transition point   (e.g., a rigorous proof that $\mathcal{C}$ is extensive for translationally invariant gapped systems in the low-temperature regime is provided in Appendix A of  \cite{Hovhannisyan2018}). The optimal controls are found to be $\lambda^*_2\approx 5, \lambda^*_1\approx 8$.
\begin{figure}
\includegraphics[width=0.6\textwidth]{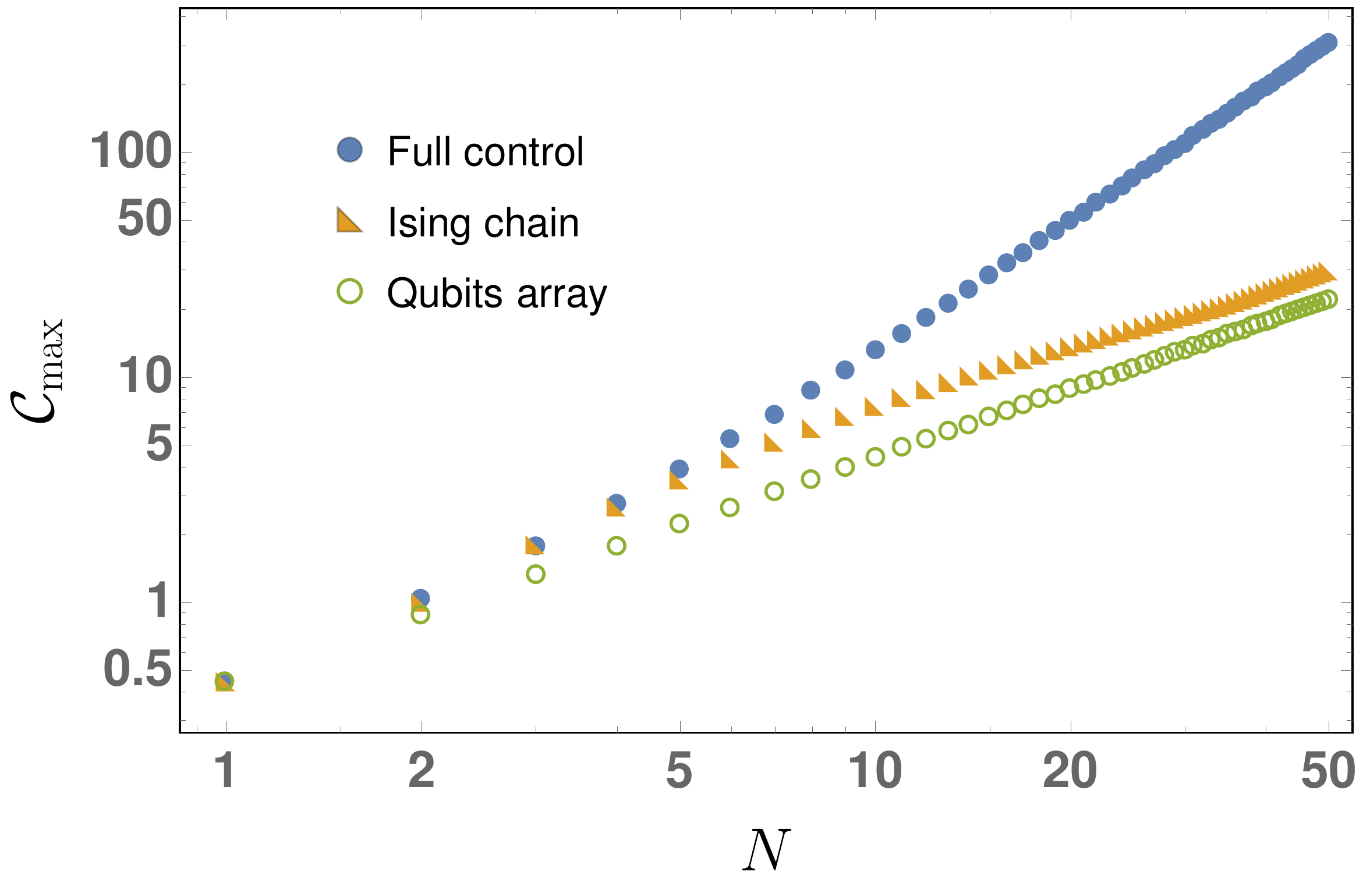}
\caption{Maximum heat capacity $\cal{C}_{\rm max}$ (corresponding to the maximum power $P^{\rm (max)}_\gamma$ in adimensional units), for a system of $N$ qubits with different degrees of control. We obtain asymptotically: ${\cal{C}}_{\rm max} \simeq 0.44 N$ for  $N$ independent qubits with a single control parameter, ${\cal{C}}_{\rm max} \simeq 0.59 N$  for an Ising chain with two control parameters, and  ${\cal{C}}_{\rm max} \simeq N^2/4$  for full control on the spectrum. %When the Hamiltonian consists in $N$ non-interacting 2-level systems the optimal power is simply $N$ times the maximum power of a single qubit. In case of an Ising chain the set of possible Hamiltonians is enlarged (thus better power) but the asymptotic scaling remains linear. In case of full control of the $2^N$-dimensional Hamiltonian the maximum power scales as $N^2$
}
\label{fig:supra-extensive}
\end{figure}

\section{Asymptotic expansions and critical scaling}
\label{app:asymptoticexp}
Inspired by Ref.~\cite{ma2018universal} we can find the best power for a given fixed efficiency, i.e. $ \eta=\gamma \eta_C=\gamma \big(1-\frac{T_c}{T_h}\big)\ $, which in the symmetric low-dissipation regime  means to fix (we identify $\Delta S\equiv \Delta_h S$)
\begin{equation}
\label{eq:relate}
\frac{Q_h+Q_c}{Q_h}=\frac{(T_h-T_c)\Delta S- \Sigma (\frac{T_h}{\tau_h}+\frac{T_c}{\tau_c})}{T_h(\Delta S - \Sigma /\tau_h)} = \gamma \bigg(1-\frac{T_c}{T_h}\bigg)\ ,
\end{equation}
which relates $\tau_c$ and $\tau_h$, after which it is possible to maximize the power 
\begin{equation}
P=\frac{Q_h+Q_c}{\tau_c+\tau_h}=\frac{(T_h-T_c)\Delta S- \Sigma (\frac{T_h}{\tau_h}+\frac{T_c}{\tau_c})}{\tau_c+\tau_h}
\end{equation} 
by enforcing $\partial_{\tau_x}P=0$ (here $\tau_x$ can be either $\tau_c(\tau_h)$ or $\tau_h(\tau_c)$ via Eq.~\eqref{eq:relate}), to obtain
\begin{align}
\label{eq:max_pgamma}
&P^{\rm (max)}_\gamma= \frac{\Delta S^2}{4 \Sigma} \frac{(T_c-T_h)^2 \gamma (1-\gamma)}{T_h (1-\gamma) + T_c \gamma}
\end{align}
by choosing
\begin{equation}
\label{eq:tau_optimal}
\tau_c=\frac{2\Sigma}{\Delta S \left(\frac{T_h}{T_c}-1\right)(1-\gamma)}\ ,
\qquad
\tau_h=\frac{2\Sigma\left( \frac{T_h}{T_c}(1-\gamma)+\gamma \right)}{\Delta S\left(\frac{T_h}{T_c}-1 \right) (1-\gamma)}\ .
\end{equation}
%Note  that from the last equations the consistency of the expansion \eqref{MM-lowdis} in the main text is guaranteed whenever the ratio between the two temperatures is close to one $T_c/T_h\sim 1$, or in the limit of high efficiency $\gamma\sim 1$, which is the case analysed in this article.
The correspondent work is
\begin{align}
&W_\gamma = \frac{\Delta S(T_h-T_c)\gamma (T_c (1+\gamma)+T_h (1-\gamma) )}{2(T_h+\gamma (T_c- T_h) )}\ .
\end{align}

\subsection{Critical scaling}
Now we consider the optimal engine described in the main text whose cycle consists of $G(s\tau_c)=(1+\epsilon g(s))G(0)$ with $s\in(0,1)$, $\epsilon \ll 1 $ but finite, and $G(\tau_c+s\tau_h)=G((1-s)\tau_c)$.  $G$ is chosen appropriately to maximize the heat capacity given the allowed control. In the limit $\epsilon\ll 1$ the shape $g(s)$ becomes irrelevant as far as it is a smooth function with $g(0)=0$ and $g(1)=1$, which implies $\langle \dot{g}\rangle=1$. This leads to
\begin{align}
&\Delta S \simeq \epsilon  \hspace{1mm} {\rm cov}_{\omega} (G,G)
\\
& \Sigma\simeq \epsilon^2 \tau_{\rm eq} \hspace{1mm} {\rm cov}_{\omega} (G,G)\ .
\end{align}
We also consider that we scale up the engine with $N$ while approaching $\gamma \rightarrow 1$, by setting (using the notation for critical exponents of a second order phase transition as in \cite{critical_engine,critical_engine_fluctuations})
\begin{align}
&1-\gamma= N^{-\xi}, \hspace{8mm} \xi>0, 
\nonumber\\
& {\rm cov}_{\omega} (G,G)= c_0 N^{1+\alpha/{(d\nu)}}, \hspace{8mm} \alpha \geq 0,
\nonumber\\
&\tau_{\rm eq}  = \tau_0 N^{{z}/{d}}.
\label{app:gamma}
\end{align}
%One can obtain $\alpha-z\nu>0$ for certain materials, see paper Campisi (\textcolor{red}{to be expanded, here or in the main text}).  
Then, expanding the relevant quantities for $N\gg 1$, we obtain at leading order in $N$:
\begin{align}
\label{eq:asymptotic_expansion}
P&=\frac{c_0(T_c-T_h)^2}{4\tau_0 T_c}N^{1+\alpha/(d\nu)-z/d-\xi}+\mathcal{O}(N^{1+\alpha/(d\nu)-z/d-2\xi})
\nonumber\\
\tau_c&= \frac{2\epsilon}{\frac{T_h}{T_c}-1}  \tau_0 N^{\xi+z/d}
\nonumber\\
\tau_h &= \tau_c +\mathcal{O}(1)
\nonumber\\
W&=\epsilon c_0 N^{1+\alpha/(d\nu)} (T_h-T_c) +\mathcal{O}(N^{1+\alpha/(d\nu)-\xi})
\nonumber\\
 &= (T_h-T_c)\Delta S +\mathcal{O}(N^{1+\alpha/(d\nu)-\xi})
\end{align}

Now we look at the fluctuations of work. The cycle consists of four processes: Two (quasi)-isothermal processes and two quenches. We note that as $N\rightarrow \infty$, we have that $W \rightarrow (T_h-T_c)\Delta S$, so that the isothermal processes become exact at leading order in $N$  which implies that they become fluctuationless  \cite{martimatteo_fluctuations}; note that this is expected as  $\tau_c \rightarrow \infty$ with $N\rightarrow \infty$, which makes the low-dissipation assumption also more and more exact as we increase $N$. Secondly, regarding the two quenches: $H \rightarrow H T_h/T_c$ and the inverse one $H \leftarrow H T_h/T_c$; it is easy to see that  the work fluctuations are given by
\begin{align}
\sigma_W^2= 2\left(T_h-T_c \right)^2 \left[ {\rm Tr}(\omega G^2)- ({\rm Tr}(\omega G)^2 \right]= 2\left(T_h-T_c \right)^2 {\rm cov}_{\omega} (G,G) \ ,
\end{align}
and thus
\begin{align}
\sigma_W^2=   2\left(T_h-T_c \right)  \frac{W}{\epsilon }.
\end{align}
Hence we have that
\begin{align}
f_w=\frac{\sigma_W}{W} \propto \frac{1}{\sqrt{\epsilon W}} \propto \frac{1}{\epsilon \sqrt{ N^{1+\alpha/(d\nu)}}}\ .
\end{align}
As explained in the main text,  if we want to exploit the critical scaling in the phase transition of the system $\epsilon$ must satisfy $\epsilon\leq N^{-1/(2-\alpha)}$. Then, using  $d\nu=2-\alpha$ \cite{huang2009introduction}, we obtain
\begin{align}
f_w\propto \frac{1}{\epsilon \sqrt{ N^{1+\alpha/d\nu}}}= \frac{1}{\epsilon N^{(1+\alpha/(2-\alpha))/2}}\geq\frac{N^{1/(2-\alpha)}}{N^{1/(2-\alpha)}}=1 \ ,
\end{align}
which is the same result found in \cite{critical_engine_fluctuations} for the Otto engine proposal. More details on the necessary critical exponents, and on the fluctuations of the engine, are provided in the main text. % Nevertheless the above ratio is referred to a single cycle, that is it considers the ratio of fluctuations to work \emph{per cycle}. Given that the fluctuations in each cycle are independent, after $M$ cycles the ratio of the total fluctuations over the total work will have the standard scaling $\sqrt{1/M}$ (\textcolor{blue}{this is essentially a consequence of the law of large numbers, so it holds for $M$ large})

\subsection{Comparison with Otto cycle}

Let us check in more detail how our results compare to Ref. \cite{critical_engine}, where an Otto cycle is considered. Following  \cite{critical_engine}, let us define the internal energy
\begin{align}
U_H(\beta)= \Tr \left(   H \frac{e^{-\beta H}}{\Tr(e^{-\beta H})}\right)
\end{align}
and the heat capacity
\begin{align}
\mathcal{C}_H (\beta)= - \beta^2 \frac{\partial U_H(\beta)}{\partial \beta}\ . 
\label{HC} 
\end{align}
The work output of a Otto cycle working between Hamiltonians $\lambda_h H \leftarrow \lambda_c H$ for a fixed efficiency  $\eta= \gamma \eta_C$, $\Delta \eta= \eta_C (1-\gamma)$ is given by 
\begin{align}
W=-\lambda_h \eta_C \gamma \left[U_H (\beta_h \lambda_h+\beta_c \lambda_h \Delta \eta )-U_H(\beta_h \lambda_h)\right] \ .
\label{Wexact}
\end{align}
Expanding for low $\Delta \eta$ which corresponds to $\lambda_c H_c \approx \lambda_h H_h$, using \eqref{HC} and keeping only leading terms in $\Delta \eta$ we obtain
\begin{align}
W\approx \eta_C^2 \frac{\beta_c}{\beta_h^2} (1-\gamma) \gamma \mathcal{C}\ ,
\label{Ottoww}
\end{align}
where we defined
\begin{align}
 \mathcal{C} \equiv \mathcal{C}_{\lambda_h H_h} (\beta_h) \approx \mathcal{C}_{\lambda_c H_c} (\beta_c) \ .
 \label{phasetransitionpoint}
\end{align}
It is important to note that the linear expansion \eqref{Ottoww} of \eqref{Wexact} is only justified close to the phase transition point \eqref{phasetransitionpoint}, whose width scales as $\delta \propto N^{-1/(d\nu)}$. This sets an extra requirement on $\Delta \eta$:
\begin{align}
\Delta \eta=\eta_C (1-\gamma) \propto N^{-1/(d\nu)}
\label{extrareq}
\end{align}
The power is simply $W/\tau$ where $\tau$ is the  time of the cycle, which involves two thermalization processes \cite{critical_engine}. Let us take $\tau= 2 \kappa \tau_{\rm eq}$, where $\kappa $ measures how exact the thermalisation process is (the error being exponentially small with $\kappa$ if one assumes a standard exponential relaxation).  Then the power reads 
\begin{align}
P_{\rm Otto}\approx \frac{1}{2\kappa \tau_{\rm eq}} \eta_C^2 \frac{\beta_h}{\beta_c^2} (1-\gamma)\gamma \mathcal{C}\ .
\end{align}
Let us now consider the Carnot cycle. After maximisation over $(\Delta S)^2/\Sigma$ we have shown in the main text that
\begin{equation}
P_\gamma^{(\rm max)}=\frac{\mathcal{C}}{4\tau_{\rm eq}}\frac{(T_h-T_c)^2\gamma(1-\gamma)}{\gamma T_c+(1-\gamma)T_h}\ .
\end{equation}
Expanding around $\gamma \approx 1$ , we have
\begin{align}
P_{\rm \gamma \approx 1}^{\rm max} \approx \frac{\kappa P_{\rm otto}}{2},
\end{align}
which is correct for a fixed $\gamma$ close to 1. 
%We however stress that $P_{\rm otto}$ must have a convergence  of $\gamma$ to 1 satisfying \eqref{extrareq}, which is not necessary in the Carnot cycle as discussed in the main text. 

\section{Explicit protocol for a $2^N-1$ degenerate system}
\label{SecAppExplicitSol}
For the sake of completeness, we show here how the results of the paper apply explicitly for a feasible driving protocol close to the optimal one, in the case
of a $D$-level system with a ground state and $D-1$ degenerate excited states. As we explained in Sec.~\ref{sec:heatcap_models} case {\bf (a)}, this Hamiltonian is motivated as it maximises the heat capacity given a $D$-level system. For comparison purposes, we will   sometimes use $N$ satisfying  $D=2^N$, so that the Hamiltonian can be thought of $N$ suitably interacting particles. 
%of full control over a spectrum \textcolor{blue}{ of dimensionality $D=2^N$.  As discussed above, this  may be compared with N two-level systems} (see Sec. \ref{sec:heatcap_models} case {\bf (a)}). %\textcolor{blue}{That is, we assume that the  $D=2^N$ energy levels of the system can be controlled at will. 
%As previously discussed, this case is interesting from a fundamental point of view, as it provides an ultimate bound on the maximal power assuming perfect control on the spectrum.  }
 Note that the case $N=1$ corresponds to the driving of a qubit \cite{cavina_optimalcontrol,abiuso_n-M}.

Consider a $D$-level system in a thermal state at temperature $1/\beta$ with an engineered Hamiltonian with a ground state and a $d\equiv D-1$ degenerate excited states with gap $E/\beta$, i.e.
\begin{equation}
H(t)=\frac{E(t)}{\beta} \Pi^{(d)}, \qquad \Pi^{(d)}=\sum_{i=1}^{d} |i\rangle\langle i|, \qquad d\equiv 2^N-1,
\end{equation}
such that the thermal state is
\begin{equation}
\omega(t)=\frac{|0\rangle\langle 0|+\Pi^{(d)}e^{-E(t)}}{1+de^{-E(t)}}\equiv q(t)|0\rangle\langle 0|+(1-q(t))\Pi^{(d)},
\end{equation}
where we defined the driving ground state population
\begin{equation}
q(t)=\frac{1}{1+de^{-E(t)}}\ .
\end{equation}
Consider the dynamics given by the equation
\begin{equation}
\frac{d}{dt}\rho(t)=\Gamma(\omega(t)-\rho(t)), \qquad \Gamma\equiv \frac{1}{\tau_{\rm eq}}
\end{equation}
(which induces the metric (4) of the main text in the context of continuous time dynamics, as explained in Section \ref{sec:STD_thermolength} or Ref.~\cite{marti2018thermodynamiclength}).
We can then consider a solution in the form
\begin{equation}
\rho(t)=p(t)|0\rangle\langle 0|+(1-p(t))\Pi^{(d)}
\end{equation}
characterized in terms of the ground state probability, which by the slow-driving approximation \cite{slowdriving} can be solved to be
\begin{equation}
p(t)=q(t)-\frac{1}{\Gamma}\frac{d}{dt}q(t)+\frac{1}{\Gamma^2}\frac{d^2}{dt^2}q(t)+\dots
\end{equation}
that is, expressing quantities in terms of the adimensional time unit $\rho(t)=\rho(s\tau)$ (here $\tau$ is the duration of the driving), and using the notation $\dot{A}=\partial_s A=\tau\frac{d}{dt}A$,
\begin{equation}
\rho=\sum_{i=0}^{\infty} \rho^{(j)}, \quad \rho^{(j)}=(-1)^{j}\frac{1}{\tau^j\Gamma^j}\partial^{(j)}_s\omega
\end{equation}
The heat exchange can be computed at all the orders
\begin{equation}
\label{eq:explicit_heat_orders}
\beta Q^{(j)}=\int_0^1 \Tr[\beta H\dot{\rho}^{(j)}] ds=\int_0^1 \sum_{i=1}^d E(-\tau\Gamma)^{-j} \partial^{(j+1)}_s\bigg(\frac{1-q}{d}\bigg) ds=-\int_0^{1} \frac{E}{(-\tau\Gamma)^{j}}\partial^{(j+1)}_s q \ ds\ .
\end{equation}
The point of maximum heat capacity that optimizes the power output of this system, as shown in Sec. \ref{sec:heatcap_models} case {\bf (a)}, is asymptotically $E\sim \ln d$ which implies $q\sim \frac{1}{2}$. We consider thus a modulation near this point, i.e. a driving protocol in the form $q(t)=\frac{1}{2}\big(1+\varepsilon\cos(\pi t/\tau)\big)$, with a small abuse of notation
\begin{equation}
\label{eq:pop_driving}
q(s)=\frac{1}{2}\big(1+\varepsilon\cos(\pi s)\big) ,
\end{equation}
or equivalently $E(s)=\ln(\frac{d q(s)}{1-q(s)})$.
Notice here that we chose to use the letter $\varepsilon$ instead of $\epsilon$ used in the main text and in Eq.~\eqref{eq:robustness_degmod}, because the latter multiplies the Hamiltonian, which scales linearly in $N$ on the optimal point, while here the modulation of $E(s)$ is of order $\varepsilon$ with no pre-factor. Hence for scaling comparisons it is important to remember $\epsilon\sim\varepsilon/N$.

\begin{figure}
\includegraphics[width=0.38\textwidth]{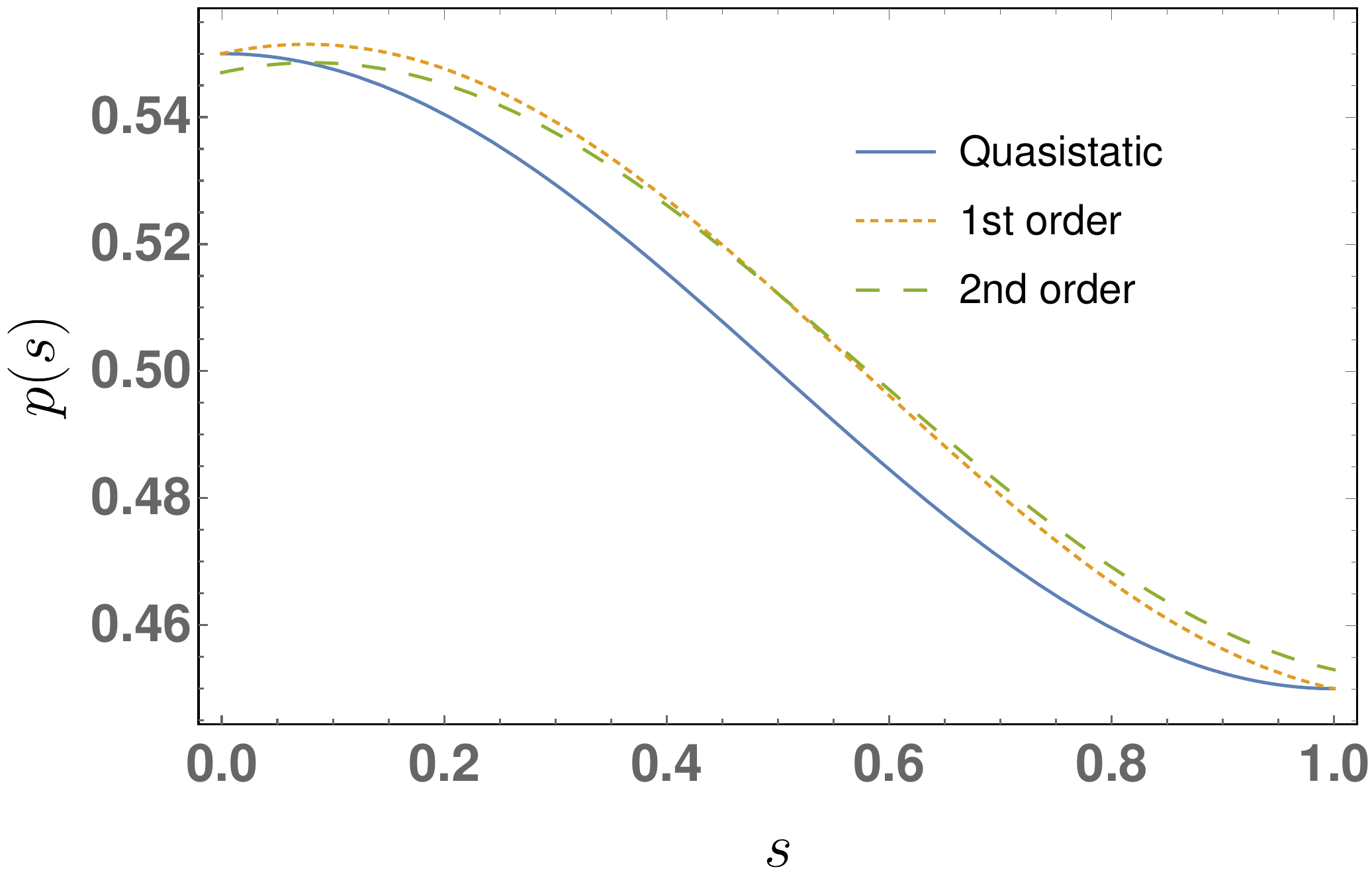}
\quad
\includegraphics[width=0.52\textwidth]{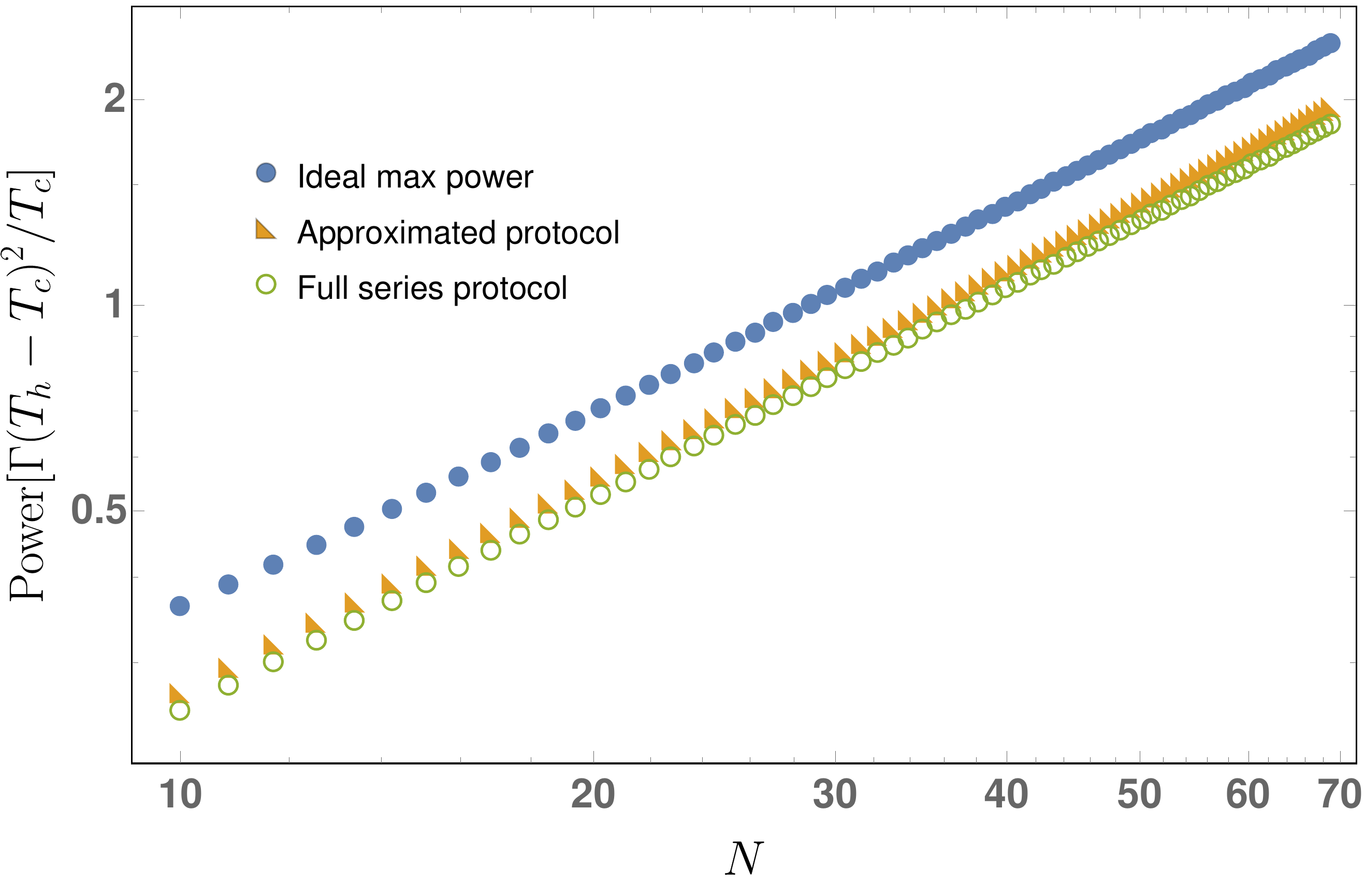}
\caption{(Left) Ground state probability evolution in adimensional units $p(s)$, for the driving given in Eq.~\eqref{eq:pop_driving} for $\epsilon=0.1$, $T_c/T_h=0.9$, $\gamma=0.9$, $N=10$. Different orders of approximation are compared.\\
 (Right) Extensive Power scaling of the heat engine as a function of its size $N$ (the number of qubits it is composed of), obtained by the above explicit feasible driving of Eq.~\eqref{eq:pop_driving}, with $\epsilon=0.1$, $T_c/T_h=0.9$, in natural units. The efficiency approximates the Carnot's one as $N$ grows  $\gamma=1-\frac{1}{N}$. The blue circles represent the maximal ideal bound, the yellow triangles represent the performance achieved by the protocol using the slow-driving approximation, while the green rings show the correction to the full analytical series.}
\label{fig:degmod}
\end{figure}

The total heat can be explicitly computed from \eqref{eq:explicit_heat_orders}
\begin{align}
\nonumber
\beta Q=\int_0^{1} -E \bigg(\dot{q}-\frac{\ddot{q}}{\tau\Gamma}+\frac{\dddot{q}}{\tau^2\Gamma^2}+\dots\bigg)=
\int_0^1 -E\frac{\varepsilon}{2}\bigg(-\pi\sin(\pi s)+\frac{\pi^2\cos(\pi s)}{\tau\Gamma}+\frac{\pi^3\sin(\pi s)}{\tau^2\Gamma^2}+\dots\bigg)\\
\int_0^1 -E\frac{-\pi\varepsilon}{2}\big(\sin(\pi s)-\frac{\pi}{\tau \Gamma}\cos(\pi s)\big)\big(1-\frac{\pi^2}{\tau^2\Gamma^2}+\frac{\pi^4}{\tau^4\Gamma^4}-\dots\big)=\frac{\pi\varepsilon}{2}\frac{1}{1+\frac{\pi^2}{\tau^2\Gamma^2}}\int_0^1 E\big(\sin(\pi s)-\frac{\pi}{\tau \Gamma}\cos(\pi s)\big)\ .
\end{align}
The low-dissipation we considered in the text consists in neglecting all the terms of order $O(\frac{1}{\tau^2\Gamma^2})$ on (which in this specific case would consist only in a renormalization of the total result), hence identifying $\beta Q=\Delta S-\Sigma/\tau$:
\begin{equation}
\Delta S_{LD} = \frac{\pi\varepsilon}{2}\int_0^1 E(s)\sin(\pi s) ds \qquad \Sigma_{LD}=\frac{\pi^2\varepsilon}{2\Gamma}\int_0^1 E(s)\cos(\pi s)\big)\ .
\end{equation}
%More precisely, given the solution for the ground state probability
%\begin{equation}
%p(s)=q(s)-\frac{\dot{q}(s)}{\tau\Gamma}+\frac{\ddot{q}(s)}{\tau^2\Gamma^{2}}-\dots
%\end{equation} 
%it is possible to compute exactly
%\begin{equation}
%\Delta S=\Delta \big(-p\ln p -(1-p)\ln(\frac{1-p}{d})\big)= -\int dp \ln\big(\frac{dp}{1-p}\big)=-(\ln d) \Delta p+\Delta H_2(p)
%\end{equation}
%where $H_2$ represents the binary Shannon entropy, while
%\begin{align}
%\beta\Delta Q=& -\int ds\dot{p} \ln\big(\frac{dq}{1-q}\big)=-(\ln d)\Delta p -\int ds\dot{p} \ln\big(\frac{q}{1-q}\big)=-(\ln d)\Delta p -\int ds\big(\dot{q}-\frac{\ddot{q}}{\tau\Gamma}+\frac{\dddot{q}}{\tau^2\Gamma^2}\big) \ln\big(\frac{q}{1-q}\big)\\
%=& -(\ln d) \Delta p+\frac{\Delta H_2(q)}{1+\frac{\pi^2}{\tau^2\Gamma^2}}+\frac{1}{1+\frac{\pi^2}{\tau^2\Gamma^2}}\int dt\frac{\ddot{q}}{\tau\Gamma} \ln\big(\frac{q}{1-q}\big)
%\end{align}
%given the symmetry of the protocol around $q=1/2$, we have $\Delta H_2(q)=0$ and $\Delta H_2(p)=0$, hence
%\begin{equation}
%\Delta S=-(\ln d) \Delta p= \frac{-(\ln d) \Delta q}{1+\frac{\pi^2}{\tau^2\Gamma^2}}=\frac{\pi\varepsilon}{2}\frac{1}{1+\frac{\pi^2}{\tau^2\Gamma^2}}\int_0^1 E\sin(\pi s)=\frac{1}{1+\frac{\pi^2}{\tau^2\Gamma^2}}\Delta S_{LD}
%\end{equation}
%and
%\begin{equation}
%\beta\Delta Q=\Delta S+\frac{1}{1+\frac{\pi^2}{\tau^2\Gamma^2}}\int dt\frac{\ddot{q}}{\tau\Gamma} \ln\big(\frac{q}{1-q}\big)=\frac{1}{1+\frac{\pi^2}{\tau^2\Gamma^2}}\bigg(\Delta S_{LD}-\frac{\Sigma_{LD}}{\tau}\bigg)
%\end{equation}
In the high efficiency regime we can then write from the asymptotic expansions of Eq.~\eqref{eq:max_pgamma} and \eqref{eq:tau_optimal}
\begin{align}
P^{\rm (max)}_\gamma\simeq \frac{\Delta S^2}{\Sigma} \frac{(T_c-T_h)^2  (1-\gamma)}{4T_c \gamma}\\
\tau_h\simeq\tau_c\simeq\frac{2\Sigma}{\Delta S \left(\frac{T_h}{T_c}-1\right)(1-\gamma)}\ .
\end{align}
In Figure \ref{fig:degmod} we show the evolution of the ground state probability $p(s)$ (which characterizes the whole state solution) according to different approximations, as well as the actual performance of the engine compared to the maximal one (which is obtained in the limit $\varepsilon\rightarrow 0$), for feasible choices of the parameters. The slow-driving approximation gives results that almost coincide with the full analytical series for this particular protocol, and the performance is close to the optimal one.

\end{document}